\documentclass[acmsmall]{acmart}

\newcommand{\yellowbox}[2]{
\vspace{10pt} 
\begingroup
\setlength{\fboxsep}{10pt} 
\setlength{\fboxrule}{2pt} 
\noindent\fcolorbox{yellow!75!black}{yellow!5!white}{%
\begin{minipage}{\dimexpr\linewidth-2\fboxsep-2\fboxrule\relax}
  \textbf{#1} 
\end{minipage}%
}
\endgroup
\vspace{10pt} 
}

\AtBeginDocument{%
  }

\setcopyright{acmlicensed}
\copyrightyear{2018}
\acmYear{2018}
\acmDOI{XXXXXXX.XXXXXXX}
\acmConference[Conference acronym 'XX]{Make sure to enter the correct
  conference title from your rights confirmation email}{June 03--05,
  2018}{Woodstock, NY}
\acmISBN{978-1-4503-XXXX-X/2018/06}

\usepackage{xspace}
\usepackage{multirow}
\newcommand{\sysname}{Magentic-UI\xspace}

\setcopyright{rightsretained}

\begin{document}

\title{Overseeing Agents Without Constant Oversight: Challenges and Opportunities}

  
\author{Madeleine Grunde-McLaughlin}
\email{mgrunde@cs.washington.edu}
\authornote{Work completed during an internship at Microsoft Research}
\affiliation{
  \institution{University of Washington}
  \country{United States}
}

\author{Hussein Mozannar}
\email{hmozannar@microsoft.com}
\affiliation{
  \institution{Microsoft Research}
  \country{United States}
}

\author{Maya Murad}
\email{mayamurad@microsoft.com}
\affiliation{
  \institution{Microsoft Research}
  \country{United States}
}

\author{Jingya Chen}
\email{jingyachen@microsoft.com}
\affiliation{
  \institution{Microsoft Research}
  \country{United States}
}

\author{Saleema Amershi}
\email{samershi@microsoft.com}
\affiliation{
  \institution{Microsoft Research}
  \country{United States}
}

\author{Adam Fourney}
\email{Adam.Fourney@microsoft.com}
\affiliation{
  \institution{Microsoft Research}
  \country{United States}
}

\renewcommand{\shortauthors}{Grunde-McLaughlin et al.}

\begin{abstract}
  To enable human oversight, agentic AI systems often provide a trace of reasoning and action steps. 
Designing traces to have an informative, but not overwhelming, level of detail remains a critical challenge. 
In three user studies on a Computer User Agent, we investigate the utility of basic action traces for verification, explore three alternatives via design probes, and test a novel interface's impact on error finding in question-answering tasks.
As expected, we find that current practices are cumbersome, limiting their efficacy. 
Conversely, our proposed design reduced the time participants spent finding errors. However, although participants reported higher levels of confidence in their decisions, their final accuracy was not meaningfully improved. 
To this end, our study surfaces challenges for human verification of agentic systems, including managing built-in assumptions, users' subjective and changing correctness criteria, and the shortcomings, yet importance, of communicating the agent's process.

\end{abstract}


\begin{teaserfigure}
  \includegraphics[width=\textwidth]{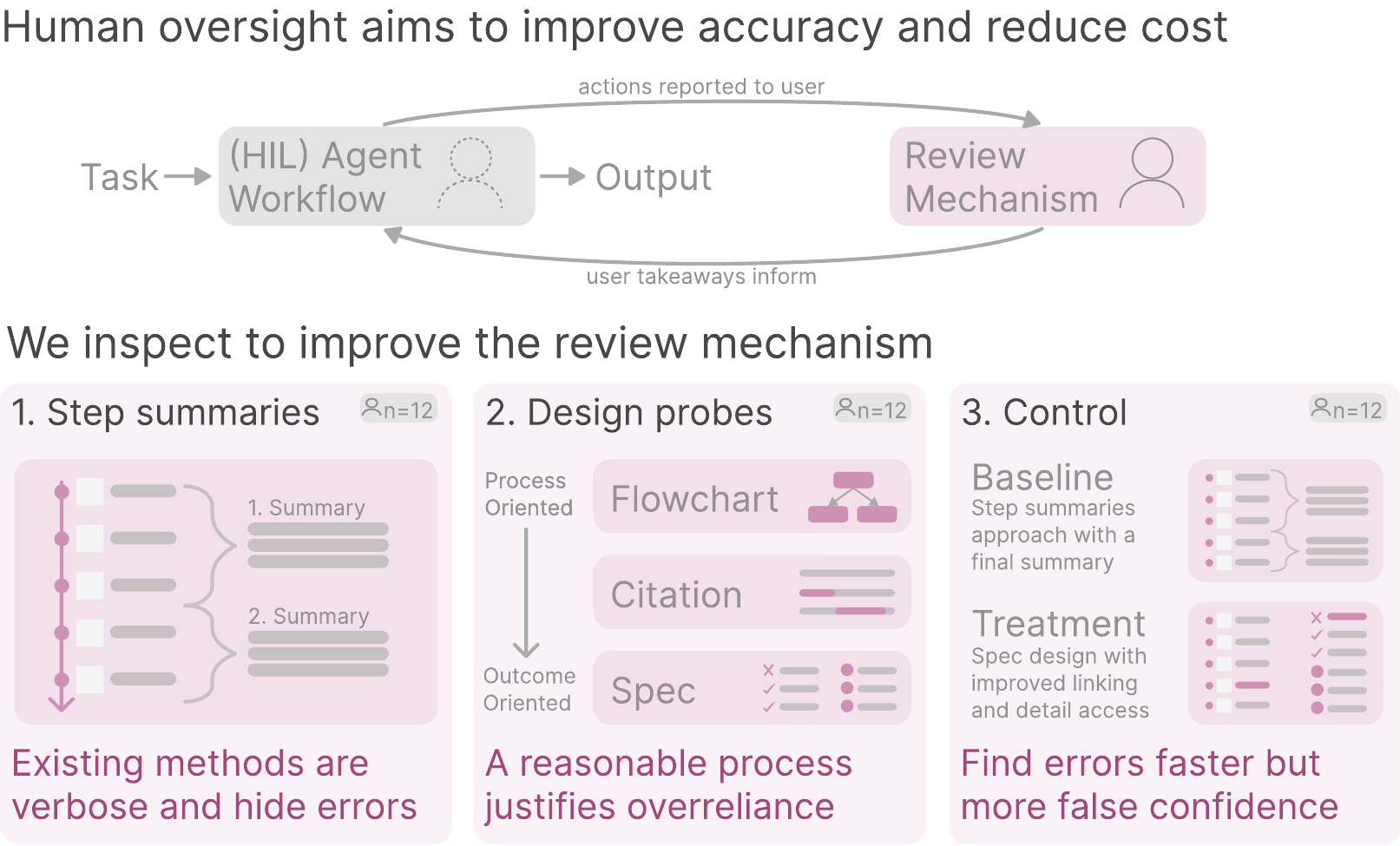}
  \caption{Through three user studies, we investigate how the presentation of an agentic workflow's actions affects human verification. We first find that participants miss small but impactful errors with the current verbose display listing agent steps with interleaved summaries. 
  With three design probes, we next find that focusing on the steps provides useful context but can justify overreliance. 
  Finally, with our novel interface that explicitly defines the task's requirements and the agent's assumptions, participants found errors faster but also reported increased confidence with no meaningful effect on accuracy. }
  \Description{There are two rows of items. One is titled "Human oversight aims to improve accuracy and reduce cost". Under that heading, there are two boxes, one of which says ``(HIL) Agent Workflow) that has a ``Task'' input and an ``Output'' output. An arrow connects this box to another titled ``Review Mechanism''. This arrow says ``actions reported to user''. There is an arrow going from the ``Review Mechanism'' box to the ``(HIL) Agent Workflow'' box that says ``user takeaways inform''. The second row of items is titled ``We inspect to improve the review mechanism.'' This section is split into three columns, each of which represent a user study and have an n=12 indication. First is titled ``Step summaries'' and has the takeaway ``Existing methods are verbose and hide errors.'' It shows a diagram of small action steps that are split into groups of actions that are summarized in text. The second column says ``Design probes'' with the takeaway ``A reasonable process justifies overreliance.'' It has three items along an arrow going from ``Process-Oriented'' to ``Outcome'' with three methods ``Flowchart'', ``Citation'' and ``Spec'' along that arrow. The third column is titled ``Control'' with the takeaway ``Find errors faster but more false confidence.'' It has two parts, the ``Baseline: Step summaries approach with a final summary'' and ``Treatment: Spec design with improved linking and detail acess''. }
  \label{fig:pull}
\end{teaserfigure}

\received{20 February 2007}
\received[revised]{12 March 2009}
\received[accepted]{5 June 2009}

\maketitle

\section{Introduction}

Automated agents, discussed as an interaction technique for decades~\cite{lieberman1997autonomous,maes1994agents,shneiderman1997direct,shoham1993agent}, have gained traction for use in complex tasks following recent advances in large language models. 
Coding agents like Claude~\cite{anthropic2024claudecode} and Github Copilot~\cite{githubcopilot} write code and use code execution engines, Deep Research tools produce detailed reports~\cite{google2025deepresearch,openai2025deepresearch}, and Computer Use Agents (CUA) like Operator~\cite{openai2025operator} and Fara-7B~\cite{awadallah2025fara} interact with the web. 
This rapidly increasing suite of publicly available agents aims to offload tedious and difficult tasks from the user to the agent~\cite{aiagentsdirectory2025ai}. 
However, these systems remain error-prone, pose privacy and safety concerns, and their success is often context dependent~\cite{chen2025toward,chan2023harms,pang2025interactive}. 

To mitigate these risks and retain human agency, an expanding body of work incorporates human-in-the-loop (HIL) features into agentic systems~\cite{zou2025survey}. 
For example, the HIL CUA \sysname enables users to edit the plan of action before it occurs, take over the browser at any point, and approve of consequential actions. 
With such control, a user performing a task like buying tickets to local museums can change the plan to prefer the museums' official website over third parties, perform navigation actions if the agent struggles, and give explicit approval before payment.

Before a user can take control to correct an error or provide clarification, they must find and define the problem. In the above scenario, if the system selects tickets for the wrong day, or if the museum is actually closed on the user-specified day, steering the output to avoid these complications requires first recognizing that they occur. 
To make the agent's actions and errors visible to the user, current systems output information about the agent's progress. Such information includes an overview of planned steps, text descriptions of actions taken and their motivations, summarizations of sets of these actions, citations to references, and screenshots of the browser if applicable~\cite{mozannar2025magentic,openai2025gpt51thinking,feng2024cocoa,google2025deepresearch}. We use the term \emph{trace} to refer to this provided information. 

To be effective, a trace should enable users to find errors accurately and without expending needless effort. Unfortunately, extensive prior work finds that explaining AI behavior to users often does not improve comprehension,  decision making, or appropriate reliance, thus necessitating careful design~\cite{bansal2021does,vaccaro2024combinations,buccinca2021trust,vasconcelos2025generation,ibrahim2025measuring}. 
Initial work suggests that such comprehension remains challenging for agentic workflows. One study found that, of all interaction types, participants spent the most time on output inspection
~\cite{feng2025regulatory}, and communication errors occurred in 65\% of cases with a real user in the Collaborative Gym benchmark~\cite{shao2024collaborative}.
With cumbersome communication, users spend less effort verifying or take longer to verify than they would to do the task themselves, reducing the utility of the overall system~\cite{vasconcelos2023explanations}.

We \textbf{investigate and improve upon agentic trace presentation} for the CUA \sysname~\cite{mozannar2025magentic} in a series of user studies (Figure~\ref{fig:pull}). 
We first ran a formative study (n=12) to identify how users verify with existing supports. We found that the current trace design is verbose, cumbersome, and time-consuming, with participants missing small but impactful errors. 
Aiming to improve user experience and error-finding capabilities, a subsequent study (n=12) tested the efficacy of three design probes that summarize the trace content: Flowchart, Citation, and Specification summary. Participants were presented with an already-concluded task and were asked to determine if the system's answer to a question was correct. Flowchart, the most process-oriented probe of the three designs, had the lowest error-finding rate and was the least preferred by participants. Specification had the highest error-finding rate and was most preferred, while Citation had middling performance.
In a final study, we incorporated our prior findings into a single novel interface. This study determined that participants (n=12) found errors faster (Hedges' $g$: -0.65) with our interface. However, we found no meaningful effect on improving overall accuracy (Hedges' $g$: 0.18). More worrisome, when errors were missed, participants reported higher confidence in the agent's correctness (Hedges' $g$: 0.85).

Our iterative design process \textbf{characterized crucial but under-served challenges for human validation of multi-step reasoning models and agentic workflows}.
For instance, when participants mistakenly thought an incorrect answer was correct, they often cited a ``reasonable'' process taken by the agent as their justification. Presentations that highlight the overall process rather than the execution of each step exacerbated this overreliance. However, participants also felt lost when the presentation was divorced from the process's context. Thus, there is a tension between communicating the overall process while grounding in execution details.
Another challenge is supporting users' review of information beyond the agent's taken actions and provided justifications, such as including searched for but unfound information, uninvestigated sources, and impactful decision points that were not significant enough to warrant an interruption to the user.
Finally, users often had a subjective and wandering sense of ``correctness'' as the CUA progressed in completing the task, such that user review is likely to remain a critical practice even as models improve.

Thus, we find that the standard approach for many HIL agentic workflows of \textbf{describing and summarizing the planning steps and actions taken does not adequately support effective human verification}, creating an illusion of accountability. Listing steps at the detail needed to surface errors is overwhelmingly verbose, but a summary of the high-level process can hide execution errors. 
By supporting trace navigation with an outcome-oriented design, we move the needle towards better trace design and elucidate challenges for future development. 

\section{Related Work}
\label{sec:related}

Human-in-the-loop agentic systems intend to provide improved performance as well as user context and agency while still reducing effort (\ref{sec:related-cua}). Human-AI interaction studies show the necessity of carefully designed interventions for verification and oversight, but existing automated methods for measuring CUA, namely benchmarks, rarely explicitly study HIL performance (\ref{sec:related-verf}). We investigate how well users interacting with a HIL CUA system are able to find errors, and we illuminate opportunities and challenges for improved oversight through iterative design.

\subsection{Human oversight of Computer Use Agents}
\label{sec:related-cua}

Improved model capabilities have powered the rapid development and deployment of AI Agents~\cite{openai2025operator,manus2025agent,anthropic2024introducing}.
The AI Agents Directory now holds 2000+ agents and is quickly expanding~\cite{aiagentsdirectory2025ai}.
Despite this rapid growth, current agents have substantial deficiencies in performance and can be costly~\cite{huq2025cowpilot,shao2024collaborative,yao2024tau,kapoor2024ai}. 

A human-in-the-loop approach, in which a person can review and steer the agent's actions throughout execution, intends to improve performance while reducing effort for the user overall~\cite{zou2025survey}. 
For instance, OpenAI's Operator~\cite{openai2025operator} allows users to take control of the browser and make manual interventions if necessary. 
Beyond error fixing, human oversight also enables incorporation of user context, preferences, and expertise~\cite{shao2024collaborative,huq2025cowpilot,song2025human,ibrahim2025measuring} as well as adaptability to changing requirements from criteria drift~\cite{shankar2024validates} and unexpected contingencies~\cite{wu2020managing}. 
Another subset of work argues for semi-autonomous systems to retain human agency~\cite{zou2025call,mitchell2025fully,bennett2023does,collins2024building,feng2025levels,chan2023harms,kulveit2025gradual}.

The HCI community has discussed and designed semi-autonomous agents for programming and interacting with the web for decades~\cite{lieberman1997autonomous,maes1994agents,shneiderman1997direct,shoham1993agent}. 
As the necessary technology was not always available, these studies often employed Wizard-of-Oz techniques~\cite{maulsby1993prototyping}.
More recent studies address human interaction with implemented HIL agentic systems, including Cocoa~\cite{feng2024cocoa}, CowPilot~\cite{huq2025cowpilot}, and Magentic-UI~\cite{mozannar2025magentic}, finding that HIL systems enable better performance than entirely autonomous approaches.
For instance, Shao et al. found that Human-Agent collaboration improved performance for both simulated users and real users, although communication challenges increased the number of steps needed to complete the task in the simulated condition~\cite{shao2024collaborative}. Another branch of emerging work visualizes task decomposition from generative AI~\cite{kazemitabaar2024improving,pang2025interactive}. 

Outside of end-user steering, prior work supports human and automated oversight of workflow execution.
Such research for multi-agent workflows helps developers debug errors to improve workflow design~\cite{wu2022ai,wu2022promptchainer,epperson2025interactive,pan2025agentcoord,dibia2024autogenstudio} and incorporates LLM-based self-verification into workflows~\cite{shankar2024validates,wu2024autogen,parameswaran2024revisiting,du2024improving}, although building effectively self-verifying workflows remains challenging~\cite{cemri2025multi,grunde2025designing}.
These approaches draw on a long history of debugging and self-verification techniques for other complex workflows, such as those in crowdsourcing~\cite{grunde2025designing,huffaker2020crowdsourced,liem2011iterative,chen2019cicero,kobayashi2018empirical,kittur2011crowdforge}.

To reap its myriad proposed benefits, HIL systems must enable people to find and effectively react to unwanted agent behavior. 
To enrich the growing space of implemented HIL agentic workflows, our work dives deep into one aspect of HIL oversight: locating and defining errors.

\subsection{Verification of HIL agentic systems}
\label{sec:related-verf}

Transparent communication is necessary for human oversight of Human-AI and Human-Agent systems~\cite{terry2023interactive,shavit2023practices,collins2024building,feng2025regulatory,amershi2019guidelines,bansal2024challenges}. 
For instance, OpenAI's ``Practices for governing agentic AI systems'' cites the ``legibility of agent activity'' as a critical practice~\cite{shavit2023practices}.
However, there remains ample room for improvement in this practice. Users of the Cocoa HIL agentic system spent the most time on output inspection~\cite{feng2024cocoa}, and 65\% of cases displayed communication errors in Collaborative Gym's real condition~\cite{shao2024collaborative}. He et al. find that user involvement in planning does not support appropriate trust~\cite{he2025plan}. 
The verbosity of generative-AI explanations and reasoning traces can hinder user comprehension, but choosing appropriate granularity is challenging~\cite{kambhamettu2025attribution,shavit2023practices,feng2025regulatory,shome2025johnny,bansal2024challenges}. 
Additionally, generative-AI explanations may not be faithful to the actual underlying processes~\cite{shavit2023practices,agarwal2024faithfulness,turpin2023language,madsen2024are}. 
Extensive work in Explainable AI emphasizes that it is non-trivial to design explanations of AI behaviors that effectively support error finding and appropriate reliance in which users accept correct and reject incorrect AI outputs~\cite{liao2021human,miller2019explanation,buccinca2021trust,vasconcelos2023explanations,bansal2021does}. Encouraging appropriate reliance remains a critical concern for generative AI as well~\cite{ibrahim2025measuring,kim2024im,bo2025rely}. 
With poorly designed HIL supports, the efficacy of Human-AI teams can suffer; across a metareview of 106 experimental studies, Human-AI teams perform worse on average than either alone~\cite{vaccaro2024combinations}.

Although it is a critical challenge, most benchmarks do not measure the ability to support effective human oversight, instead measuring if autonomous agents achieve a given output~\cite{mialon2023gaia,liu2024visualwebbench,yao2022webshop,shi2017world,deng2023mind2web,pan2024webcanvas,zhou2023webarena,yoran2024assistantbench,xie2024osworld,jimenez2023swe}. 
These benchmarks can be useful metrics for organized evaluation and comparison, but prior work raises concerns about a sole focus on accuracy rather than cost~\cite{kapoor2024ai}, lack of fidelity to human evaluations~\cite{chang2025chatbench,shankar2024validates}, and validity for measurement in an open-ended domain~\cite{wallach2025position,salaudeen2025measurement,kapoor2024ai}.
Some benchmarks do incorporate simulated~\cite{shao2024collaborative,drouin2024workarena,yao2024tau,barres2025tau2} or real~\cite{shao2024collaborative} user interaction.
Outside of benchmarking, other studies evaluate Human-Agent collaborative techniques with user studies~\cite{huq2025cowpilot,mozannar2025magentic,feng2024cocoa}.
A growing body of work reflects on the type of content to be shown to the user, such as including both the outcome and the process~\cite{shao2024collaborative,terry2023interactive,balepur2025good,he2025plan} and explicitly encouraging ``users to cognitively engage with assumptions made in the reasoning chain''~\cite{pang2025interactive}. 

Thus, despite well-established findings that designing for effective human oversight is extremely challenging, relatively few evaluations of agentic systems explicitly study HIL methods. In our work, we investigate the assumption that current traces are sufficiently transparent to support human oversight and iterate on methods for communicating errors to users.

\section{Formative study}
\label{sec:form}

\yellowbox{Takeaway: Current review mechanisms are verbose, leaving Human-Agent teams error-prone. However, human oversight will remain critical for performance, as users display subjective and changing correctness criteria.}

Human oversight and verification of agentic workflows is an important feature for safety, agency, and performance~\cite{collins2024building,chan2023harms}. However, initial work finds that communication errors are prevalent~\cite{shao2024collaborative} and that users spend the most time on output reviewing~\cite{feng2024cocoa}. This formative study aims to set a foundation of existing verification practices and pain points, particularly when multitasking.
This section first reviews the CUA system, \sysname~\cite{mozannar2025magentic}, that we use (\ref{sec:form-system}), our study procedure (\ref{sec:form-procedure}), our findings (\ref{sec:form-results}), and takeaways to inform our design probes (\ref{sec:form-implications}).

\subsection{System}
\label{sec:form-system}
\sysname is an interface for HIL oversight of CUA that can interact with the browser, a code execution environment, and a file system in a contained Docker environment. It incorporates several HIL features, including: co-planning (users can review and edit the plan before it executes), co-tasking (users can take over the live browser), action approval (users must approve before the system takes an irreversable action), multi-tasking (multiple runs can co-occur), and final answer verification (a process summary with the final answer). 

\begin{figure*}[t]
 \centering 
 \includegraphics[width=\linewidth]{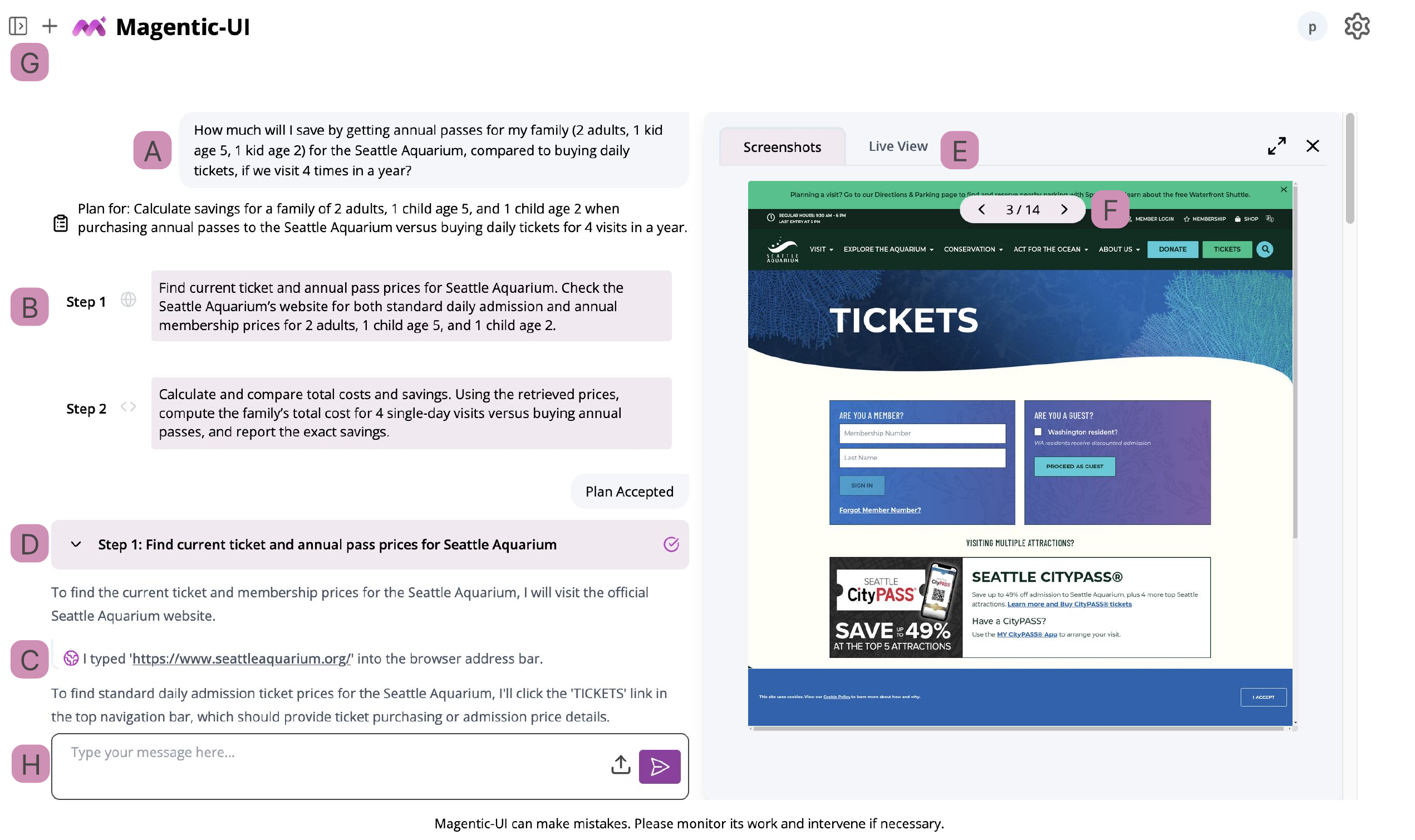}
 \caption{\sysname interface. Our formative study illuminates verification practices with the HIL CUA  \sysname~\cite{mozannar2025magentic}. Users first input a task [A], then use various features to co-plan, co-task, and adapt to errors as the system runs. To verify, they can first review the plan summary that decomposes the task into proposed plan steps [B]. Each plan step is then executed, and each action is described in the trace for transparency [C]. Collapsing these plan steps [D] shows summaries of each group of actions. While the task runs, users may watch the live view as the agent interacts with the browser [E]. This visual trace is stored in a series of screenshots as the agent progresses [F].
 Clicking the icon next to a particular action [C] links to the associated screenshot [F] to improve navigation.
 Users can open tabs between different tasks to multitask [G] and interact with chat [H]. A ``Final Answer'' summary appears upon completion (not visible).}
 \Description{An interface display of Magentic UI. The top left contains a small icon that, when clicked, opens different tabs next to the Magentic-UI logo. The left half of the screen mostly contains first the task, then a given plan of steps in distinct boxes, then a series of bullet points that show certain actions, such as ``I typed <url> into the browser address bar''. Below is a chat input box. On the right side, there are two tabs: Screenshots and Live View. The screenshots tab is selected and shows one screenshot. There is an indicator that it is showing 3/14 screenshots, with arrow buttons to the left and right. The bottom of the interface stats ``Magentic-UI can make mistakes. Please monitor its work and intervene if necessary.''}
\label{fig:formative}  
\end{figure*}

As shown in Figure~\ref{fig:formative}, the interface displays a text trace describing the task [A], the CUA's planning steps [B], each individual action [C], and mid-level summaries [D]. After execution, a ``Final Answer'' summary appears here. The other panel displays an interactive live view while the CUA uses the browser [E] and screenshots taken at regular intervals [F]. Relevant action text [C] contains an icon that, when pressed, navigates to the associated screenshot [F]. Users can move between multiple ongoing and previous tasks [G] and interact through chat [H].

\subsection{Procedure}
\label{sec:form-procedure}

Our studies were approved by our internal institutional review board (IRB). The study took place in person with \sysname pre-installed. Our twelve participants were all familiar with agentic workflows, employees of a large tech firm, and had not previously used \sysname. Participants signed a consent form, then viewed a demonstration of \sysname features. Participants were then given the opportunity to explore features by running a task directed towards finding publications. 

Next, participants completed three tasks simultaneously to encourage multi-tasking behaviors. Inspired by tasks on AssistantBench~\cite{yoran2024assistantbench}, we created three tasks that sometimes caused errors and often triggered \sysname's HIL features. These tasks were:

\vspace{0.5em}
\noindent Task 1: \textit{``Add a Hawaiian Pizza from TangleTown to my cart.'' }

\noindent Task 2: \textit{``Find all the appetizers with tuna from JustOneCookbook and save them to an alphabetized csv file with the recipe name and url.''}

\noindent Task 3: \textit{``How much will I save by getting annual passes for my family (2 adults, 1 kid age 5, 1 kid age 2) for the Seattle Aquarium, compared to buying daily tickets, if we visit 4 times in a year?''}
\vspace{0.5em}

Participants then ran a task of their own choosing. 
Next, participants completed the System Usability Scale~\cite{brooke1996sus} and a questionnaire about their familiarity with agentic systems. Finally, participants took part in a semi-structured interview. Each study was one hour long. 

Participants were encouraged to think aloud throughout the study. By hand, we extracted salient quotes from participants' think-aloud statements and semi-structured interview responses, then performed five rounds of inductive and deductive thematic analysis to organize them into themes~\cite{braun2006using}. We measured whether the Human-Agent team completed the task correctly, but we did not constrain users from making edits to the plan or from trying multiple times. We adjusted the correctness criteria according to the user's changes. Some users pivoted the plan after the initial ``Final Answer'' was output. In those instances, we counted the task completion for each separately.
For each user and task, we manually checked their results to see if the goal was accomplished. If a link was malformed, we did not count that run as incorrect if the remaining content was correct. We report a subset of results as this study informed multiple research projects.

\subsection{Results}
\label{sec:form-results}

We find that although participants wanted multi-tasking (\ref{sec:form-results-multitasking}), the existing reviewing techniques are verbose and cumbersome (\ref{sec:form-results-verbosity}). Beyond this user experience deficit, we also find that participants make small but impactful errors (\ref{sec:form-results-errors}) and have varied validation strategies (\ref{sec:form-results-strategies}). We observe that the definition of an error output can be subjective and that correctness criteria often changed as the user interacted with the system (\ref{sec:form-results-correctness}). 

\subsubsection{Participants envision multi-tasking while running CUA}
\label{sec:form-results-multitasking}

Due in large part to latency, a major pain point in the system, participants expressed a desire to multitask by leaving the system running while performing other tasks and then returning to review [P2, P4, P5, P6, P9, P10, P11, P12]. By far the most common anticipated use case of CUA was for information gathering, in which the system searches for and aggregates information that the user can then review [P1, P4, P5, P6, P7, P8, P9, P10, P11]. P8 requested that the system \textit{``curate all the information. So I'm the pilot, and it will become co-pilot,''} while P6 reflected that these tasks would be challenging to verify at scale.

\subsubsection{Verbosity makes returning to a CUA run cumbersome to review}
\label{sec:form-results-verbosity}

Participants appreciated summaries and trace provenance for understanding the state of the system when returning to an ongoing task [P4, P7, P11, P12]. 
P11 contemplated that \textit{``when you're using these AI tools, you sometimes disassociate yourself while it's like working. But then you're like, oh, wait a second, what actually happened? ... I think like having screenshots, it's just like a very nice way of like going through like the key points of like what just happened.''} However, participants noted that the reasoning text, screenshots, and plans given could be frustratingly verbose, repetitive, and not tied to discrete semantic meaning [P1, P2, P3, P5, P6, P7, P8, P11, P12]. For instance, P8 reflected that \textit{``if I don't check it immediately or if I kind of missed the point, then it's hard for me to go back.''}

\subsubsection{Participants missed small but impactful errors}
\label{sec:form-results-errors}

Of the tasks that did not run out of time or into technical errors, the Human-Agent team incorrectly completed one of nine attempts for Task 1, five of ten attempts for Task 2, and ten of twelve attempts for Task 3.
Task 1, adding a pizza to the cart, had the highest rate of success. In the one incorrect attempt, P5 added garlic puffs and fries to the order. The logs said that three items were added to the cart, which the participant read and believed. However, in the browser, only two items had been added. 
For Task 2, only one result was correct. Five were incorrect, while three, although different from the original answer and not satisfying the qualifier ``all,'' returned reasonable answers through a path we did not anticipate. One participant noticed an error but thought it was acceptable. In most incorrect instances, the CUA did not scroll to consult the appetizer recipe section examples, although the header of that section was visible. 
For Task 3, out of twelve completed attempts, ten were incorrect. Most commonly, the CUA estimated the youth price when it could not find it (n=6), although only half of the participants noticed (n=3), and those who noticed did not determine the correct answer.

We determined correctness generously, assuming that default choices were correct unless a difference was explicitly specified by the participant. However, alternative choices would have impacted the result. For instance, there were different prices every day at the aquarium, and the three participants who noticed wanted to apply different calculations (average, max, weekend-only) [P1, P3, P4]. Only two participants noticed that the CUA assumed you were not a Washington resident, one of whom was prompted by \sysname and one of whom happened to be watching the task at the right time [P5, P6]. \sysname did not successfully fulfill both participants' resulting request for resident prices.

\subsubsection{Participants employ varied verification strategies, preferring error indications in the final output, trial and error testing, and visual review}
\label{sec:form-results-strategies}

Participants ranged from cursory [P10] to thorough [P8] in the extent of their verification. Many participants focused their review on the final result rather than the trace [P5, P6, P7, P11] or validated only if they noticed something that seemed unlikely or incorrect in the output [P1, P3, P8, P10]. P1 described this process of drilling down when seeing signs of inaccuracy: \textit{``I had this like initial high level [summary] inspection available and I can click into it if I see signs of inaccuracy and ... triangulate with information from screenshots.''} Several participants raised concerns about Magentic-UI's process but decided not to verify [P5, P9, P12]. In Task 3, P9 noticed that the CUA was \textit{``not on the official website, which kinda defeats the point. Um, but anyways, I'm just gonna trust it.''} Participants would give the system requests they did not know if it could complete. They imagined that through this process, with more time, they could build a mental model of system capabilities to better know how to use HIL features and focus validation efforts  [P1, P4, P5, P6, P7, P9, P10, P11].

Participants used many elements of the interface to validate: text [P4, P5, P6, P8, P10, P11], screenshots [P1, P2, P5, P6, P8, P10, P11], code [P4, P7, P8], live view [P3, P4, P6, P7, P8, P9, P10], co-tasking [P6, P8, P9, P10], and chat [P9, P10]. Although P12 preferred reviewing text, and P2 noted that text-based explanations could be useful for screen readers, more participants expressed a preference for the visual elements [P2, P6, P8, P11]. Several participants requested more support for visual tools, including a short explanation of the current action by the cursor [P2], a flow diagram of steps [P8], and highlighted or annotated elements on the screenshots [P1, P2, P6, P7, P11].

\subsubsection{Participants display subjective and changing correctness criteria}
\label{sec:form-results-correctness}

Participants often inserted subjective preference during the planning stage [P4, P5, P6, P7, P8, P9, P11]. They also appreciated when \sysname asked clarifying questions due to underspecificaton and expressed a desire for more such questions [P2, P3, P6, P7, P11, P12], especially if consolidated during one interruption [P2, P6]. For P3's individual task of getting restaurant recommendations, they `\textit{`would like if it asked me a few questions, maybe like, you know, really basic ones like what's your budget and maybe like are you vegan.''}
When reviewing outputs, some participants also had a subjective sense of what was good enough and did not see the need to change minor inaccuracies [P1, P4, P7]. For instance, Task 2 for P1 only returned five of the six relevant recipes, which they noticed but reflected that \textit{``I appreciate the fact that it got me 5 tuna recipes, which is probably good enough.''}

However, the initial task and plan were not always sufficiently specified, as the CUA met unpredicted roadblocks and participants updated their goals. 
Participants liked when the CUA adapted to errors by adapting the plan [P5, P6], but requested a clearer delta shown from the previous plan to review updates [P8, P10, P11]. Several participants updated their goal post after seeing partial or full completion of the task, such as adding desired menu items after seeing the options [P4, P5, P7, P10, P12], changing the formatting of the originally requested output [P7, P10, P11] or creating a comparison table across multiple vendors when technical issues arose with the main restaurant site [P8]. 
Participants valued control and easy expression of their will through co-tasking [P3, P4, P5], including for following threads that were unexpected but piqued their curiosity [P8]. When multi-tasking and returning, they wished for an ability to return to a specific step with an error, make a change, then continue execution from there [P1, P6, P8, P12]. 
When called back to a trace by an action guard, they wanted to be able to give an alternative option beyond accepting or rejecting \sysname's suggestion [P5, P7, P11, P12].

\subsection{Implications from formative study informing design probes}
\label{sec:form-implications}

This formative study highlighted pain points in existing verification support, as participants missed small but impactful errors. 
Several key takeaways most inform our next study's design probes that aim to improve verification (\ref{sec:probes-designs}).
We created design probes of summarizing techniques that reduce the verbosity pain point, while emphasizing preferred visual explanations. 
Additionally, they reflect our finding that participants could not always completely pre-specify a task, such that the CUA assumed one path of actions among many during execution. 
Soliciting every small choice from the user would be overburdening, but small choices can matter, so a new design should surface elements that had been underspecified more explicitly. 
Finally, we explore the impact of focusing on the process versus the outcome, as that design dimension remained unclear.

\begin{figure*}[t]
 \centering 
 \includegraphics[width=\linewidth]{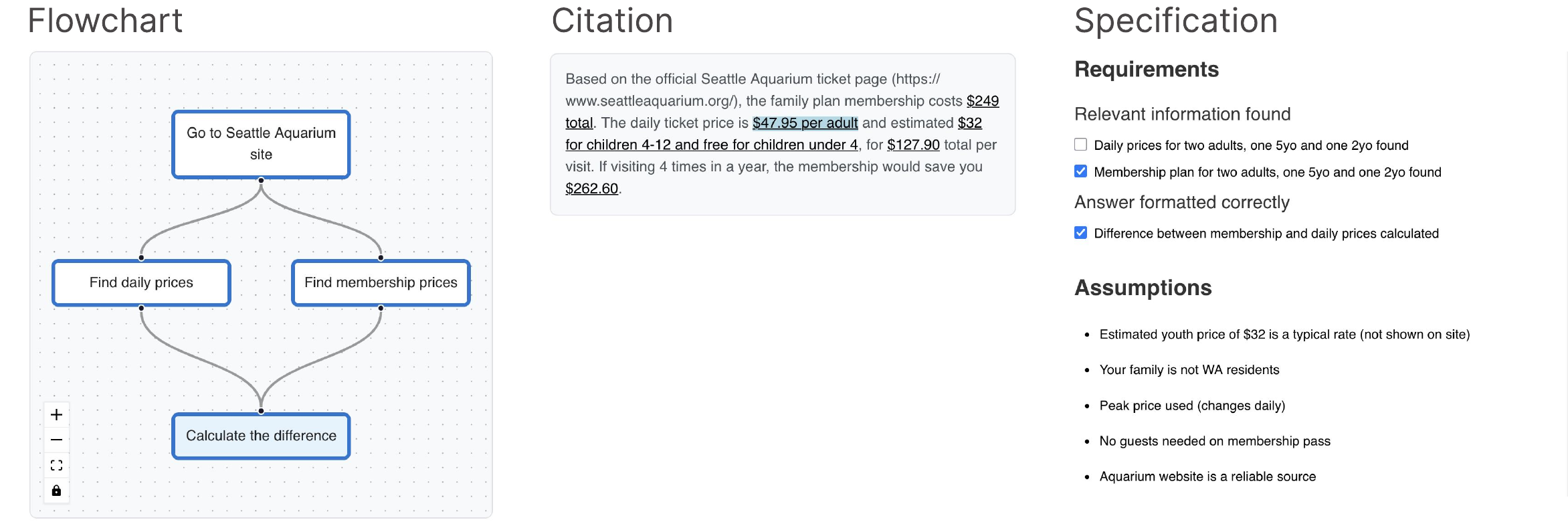}
 \caption{Three design probes. Our design probes focus on the process (Flowchart), outcome (Specification), or both (Citation). This figure illustrates each design for the task ``How much would I save by getting annual passes for my family (2 adults, 1 kid age 5, 1 kid age 2) for the Seattle Aquarium, compared to buying daily tickets, if we visit 4 times in a year?'' Additional examples are in Supplementary Materials.}
 \Description{Three columns, each of which has a method. The Flowchart column has three rows of nodes. The first node says ``Go to Seattle Aquarium site'' which connects to two nodes in the same row ``Find daily prices'' and ``Find membership prices''. Each of those connects to a node in the final row of ``Calculate the difference''. The second Citation column says the following, with [brackets] indicating underlined portions ``Based on the official Seattle Aquarium ticket page (https://www.seattleaquarium.org/), the family plan membership costs [\$249] total. The daily ticket price is [\$47.95 per adult] and estimated [\$32 for children 4-12 and free for children under 4], for [\$127.90] total per visit. If visiting 4 times in a year, the membership would save you [\$262.60].'' The final column is the specification, which includes the requirements 1) ``Daily prices for two adults, one 5yo and one 2yo found'', 2) ``Membership plan for two adults, one 5yo and one 2yo found'', and 3) ``Difference between membership and daily prices calculated'' and the assumptions ``Estimated youth price of \$32 is a typical rate (not shown on site)'', ``Your family is not WA residents'', ``Peak price used (changes daily)'', ``No guests needed on membership pass'', and ``Aquarium website is a reliable source''.}
\label{fig:designprobes} 
\end{figure*}

\begin{figure*}[t]
 \centering 
 \includegraphics[width=\linewidth]{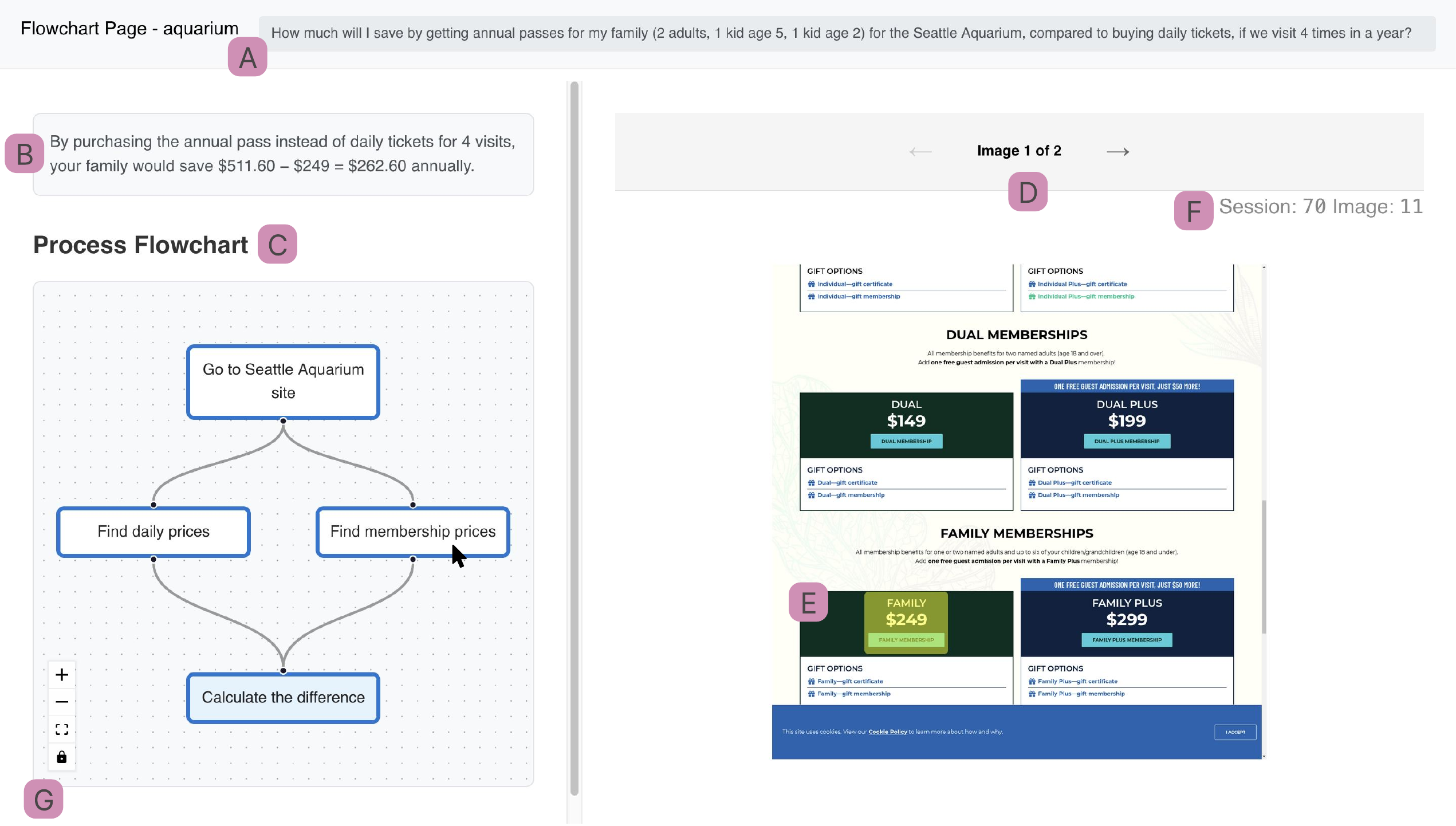}
 \caption{Design probe study interface. The interface displays the given task [A], task output [B], and one of the three design probes [C]. Clicking on the design probe surfaces the relevant screenshot(s) [D] that have highlighted annotations of the relevant information [E]. The original \sysname trace is available for further review [F]. Participants then answer the questions ``Is this result correct?'' and ``Why or why not?'' below [G].} 
 \Description{This figure shows an interface display. The top row says ``Flowchart Page - aquarium'' with the task ``How much will I save by getting annual passes for my family (2 adults, 1 kid age 5, 1 kid age 2) for the Seattle Aquarium, compared to buying daily tickets, if we visit 4 times in a year?'' The left side shows the outcome ``By purchasing the annual pass instead of daily tickets for 4 visits, your family would save \$511.60 − \$249 = \$262.60 annually'' as well as the flowchart from the design probes figure. The right side shows a screenshot with a yellow highlighted annotation over the membership price used. There is an indication that this is ``Image 1 of 2'' as well as text saying ``Session: 70 Image: 11''}
\label{fig:probeinterface} 
\end{figure*}

\section{Design probes}
\label{sec:probes}

\yellowbox{Takeaway: Abstracting to process steps reduces verbosity but justifies overreliance. An improved abstraction would provide an overview with easy detail navigation and interpretation beyond the agent's action sequences.}

We aim to build an interface that improves users' error-finding rate while minimizing the time needed for review.
With findings from the formative study (\ref{sec:form-implications}), we design three new methods for summarizing traces (\ref{sec:probes-designs}). We then conduct an exploratory, think-aloud, Wizard-of-Oz (WoZ)~\cite{maulsby1993prototyping} study (\ref{sec:probes-procedure}) and analyze the user's accuracy and length of time needed for review with each method (\ref{sec:probes-results}). Finally, we summarize the takeaway findings that inform our next controlled user study (\ref{sec:probes-implications}).

\subsection{Three design probes} 
\label{sec:probes-designs}

Both process-oriented and outcome-oriented information are valuable to communicate in the trace~\cite{shao2024collaborative,terry2023interactive,balepur2025good,he2025plan}. The three designs we test are as follows:

\vspace{0.5em}
\noindent \textbf{Flowchart.} In the Flowchart design, which most highlights the process, nodes are salient actions and edges show sequence. Boxes surrounding multiple nodes indicate a repeated task, such as finding the TripAdvisor reviews for six different hikes. Clicking on nodes with a blue outline surfaces associated annotated screenshot(s). Output nodes have a blue background.

\vspace{0.5em}
\noindent \textbf{Citation.} The Citation design has a text description of the output with justification and a brief overview of the process. Proper nouns and descriptions of CUA actions are underlined, and clicking on these underlined portions of text surfaces relevant annotated screenshot(s). 

\vspace{0.5em}
\noindent \textbf{Specification (Spec).} The Spec design is the most divorced from the process and instead reviews if the outcome meets all needs. This design has two parts: requirements and assumptions. Requirements can be determined before the task is run to create a checklist of what must be true for the output to be correct. For example, if you are checking a train ticket price for a Friday, the search should be filtered to options on Friday. Requirements are checked if completed and unchecked if uncompleted or incorrectly completed. 
Assumptions are a bullet point list of decisions that affect the output but are challenging to pre-determine. For instance, in estimating the price of museum memberships, \sysname's output assumes no guests are desired. Clicking requirements or assumptions updates the screenshots shown. Examples of each method can be found in Figure~\ref{fig:designprobes} as well as in Supplementary Materials.

All three methods show the screenshots that most saliently support important parts of the summary with highlighted yellow annotations. Arrow keys and buttons enable tabbing through multiple screenshots if needed. These may be screenshots of the browser or of the explanatory text. If \sysname captured the screenshot before the page loaded or cited text without scrolling down to it, we manually capture screenshots from the website such that the information the CUA accessed can also be accessed visually. 
On the display (Figure~\ref{fig:probeinterface}), participants view a top row containing the task [A]. On the leftmost side, they see the output answer [B], summarization method [C], and response inputs (not visible) [G]. The middle panel contains the associated screenshots [D] with the annotations [E]. The rightmost side shows the full \sysname trace (Figure~\ref{fig:formative}) [F].

\subsection{Procedure}
\label{sec:probes-procedure}

We recruited twelve participants who were employees of a large tech firm with some experience using agents. Eight participants had previously participated in the first study so were familiar with \sysname. The study was in-person with the system pre-installed. 
Participants first signed a consent form, then watched a brief demonstration of \sysname's capabilities. Participants were told to imagine that they had input these tasks and were returning to a completed task. Their job is to determine if they think \sysname's output is correct. 

We used tasks from AssistantBench, updating answers if they had changed since the benchmark's creation and preferring tasks that could not be validated from a single screenshot. To scope this initial inquiry, we filter to questions that have a correct or incorrect answer. As this was an exploratory study, each method had unique tasks, two of which were incorrect and one of which was correct. These tasks are provided in Supplementary Materials.
We used runs of \sysname using Open AI's GPT-4o~\cite{openai2024gpt4o}, the full trace of which was always available to the user on the original \sysname interface (Figure~\ref{fig:formative}). As this is a Wizard-of-Oz study, we hand-created the annotations. We follow the trace description over the ground truth. If \sysname did not complete a requirement in reality, but its trace says it is completed, we mark it as completed. 

The study is within-participant. The order of the methods and tasks was randomized to avoid ordering effects. For each method, participants were first shown and encouraged to interact with a demonstrative example. For each task, they begin by reviewing the task and answering  ``How long do you think it would take you to complete this task manually in minutes?'' with answer options from 1 to 10+. Participants then interact with the associated method. To the question ``Is the output correct?'' participants could choose between ``Yes,'' ``No,'' and ``Insufficient information to confirm.'' They are then prompted to write in free text ``Why or why not?''
We encouraged participants to think aloud throughout this process. Finally, we perform a semi-structured interview for the remaining time, approximately ten minutes. The study lasted one hour. 

We categorize participants' input responses, think-aloud comments, and semi-structured interview responses into themes using line-by-line coding. We then perform two more rounds of inductive and deductive thematic analysis~\cite{braun2006using}. We measure participants' estimated time, correctness determination, and time spent per task, including thinking aloud and responding to the input boxes.  We initially intended to classify results solely as correct or incorrect. However, as the study ran, we realized that for several examples ``Insufficient'' was a valid interpretation. Before performing the analysis, the first author determined for which tasks ``Insufficient'' is a correct answer, as marked in the Supplementary Materials.

\begin{table*}[t]
\caption{Design probe accuracy and duration performance.
The Spec method shows the highest accuracy but also the highest duration. 
GT Ans. indicates if the output is correct, with ``All'' including all trials. 
This exploratory investigation had two incorrect questions and one correct question per method.
}
\label{tab:s2-accdur}
\centering
\resizebox{0.4\linewidth}{!}{
\begin{tabular}{llrr}
GT Ans.              & Method    & Accuracy (\%) & Duration (s) \\ \hline
\multirow{3}{*}{No}  & Flowchart & 39.13         & 427.38       \\
                     & Citation  & 52.17         & \textbf{293.97}       \\
                     & Spec      & \textbf{65.22}         & 542.15       \\ \hline
\multirow{3}{*}{Yes} & Flowchart & 91.67         & \textbf{319.85}       \\
                     & Citation  & 83.33         & 573.87       \\
                     & Spec      & \textbf{100.00}          & 491.65       \\ \hline
\multirow{3}{*}{All} & Flowchart & 57.14         & 390.51       \\
                     & Citation  & 62.86         & \textbf{389.94}       \\
                     & Spec      & \textbf{77.14}         & 524.84      
\end{tabular}
}
\end{table*}

\subsection{Results}
\label{sec:probes-results}

Although participants relied on reviewing the CUA's process (\ref{sec:probes-results-useprocess}), a ``reasonable'' process could instill false confidence (\ref{sec:probes-results-overreliance}). Participants wanted additional interpretation, like highlighting alternative paths and specifying between unknown and contradictory information (\ref{sec:probes-results-alternatives}). They preferred a bird's-eye view with the option to drill down (\ref{sec:probes-results-birdseye}) but varied widely on which method most intuitively supported this summarization (\ref{sec:probes-results-preferences}).

\subsubsection{Participants utilize the process to validate and struggle when divorced from that context}
\label{sec:probes-results-useprocess}

Many participants referenced the reasonability of the process to justify trusting a correct answer [P1, P2, P3, P8, P12] or to find errors and insufficiencies [P2, P3, P5, P7, P9, P12]. For instance, P3 correctly found an error and wrote \textit{``it's clear that the flow of making decisions was missing a criteria, which is reading the specific review and see if it mentions that the reviewer has a kid.''}
If the process did not match what they thought they themselves would do, that also raised suspicions. 
P1 explains \textit{``[I] know the path it takes, and if it's different from what I'd do, [I] can double check.''} 
All participants in at least one task found the process insufficient to determine if the answer was correct or incorrect. P4 and P12 suggested live view animation as an intuitive process display.  

In the Spec and Citation designs, distance from the process could be confusing. 
P2 noted for Spec that the \textit{``context is very torn apart''} and P3 expressed that the Spec \textit{``feels too discrete and not visual enough. I don't know how to fit that understanding into my decision process.''}
For Citation, P3, P5, and P6 found it challenging to orient within the process, with P3 sharing \textit{``If I see an error early on, I immediately know everything else is not reliable. But for Citation, that's especially hard to validate if it reverses the order of steps.''}
The Spec design took longer than the other two conditions (Table~\ref{tab:s2-accdur}) and, to understand the process, participants more often referred to the original trace: eight times for Spec compared to four times for Flowchart and twice for Citation. 

\subsubsection{Justifications of overreliance referred to a reasonable process or the presence of explanations}
\label{sec:probes-results-overreliance}

In the most salient trend of overreliance, participants explained that they thought an incorrect outcome was correct because the process seemed reasonable [P2, P4, P7, P8, P10, P11]. 
For instance, P4 wrote \textit{``I think the overall process looks quite reasonable. I think I could follow a similar process as it did,''} P2 wrote \textit{``The way it started the trajectory and visited the website and the directory of the people seems logical,''} and P7 wrote \textit{``the steps it followed from the plan seem logical, so I would believe it is correct.''}
P5 reflected that the Flowchart was insufficient to find errors \textit{``unless there is like a step where the overall assumption is incorrect.''}
For incorrect tasks, participants achieved the lowest accuracy of 39.13\% on the most process-oriented condition (Flowchart) and the highest of 65.22\% on the most outcome-oriented condition (Spec), although the study was not task-balanced (Table~\ref{tab:s2-accdur}).

The existence of evidence also justified overreliance [P3, P4, P8, P11]. 
P11 wrote that \textit{``I can see the relevant pieces of information that I would've relied on to confirm/make decisions''} when incorrectly classifying an incorrect result as correct. 
Similarly, P3 stated \textit{``it shows supporting evidence for all the criteria that I can think of.''}
Finally, sometimes overreliance occurred because participants missed salient information [P1, P10]. 
For instance, participants did not view all screenshots when there were multiple supports for one piece of evidence [P1, P11, P5, P7, P2] or would believe they checked thoroughly while missing information [P3, P5, P5, P7, P8, P9, P12].

\subsubsection{Participants want additional interpretation, including highlighting alternative paths, unfound information, and sources used}
\label{sec:probes-results-alternatives}

Participants highly praised assumptions [P1, P3, P4, P5, P9, P10] and used them to find errors [P3, P5]. 
P5 asked for assumptions to be more visibly salient, and  
P1 shared that they \textit{``especially like the assumptions [showing] the other options that are possible [because when agentic browsers] give us citations, it will force us to think in the same way or say force us to go on the same path they're orienting, so it's hard for us. We can only validate by the path they go through, but we cannot validate by other sources. This would be great if there was only one source of truth, but there's so many different sources on the web.''} P12 suggested surfacing multiple options to the user together, such that \textit{``If there are multiple options for the same result, [to] do the comparison automatically, it would be wonderful. You might not know all the options that are available, or you might not compare and pick what you see first.''}

They also wanted more support for interpreting when and why the agent did not find requested information [P1, P3, P5, P9, P11, P12]. 
P9, when ranking methods, reflected that \textit{``Spec would be on top because, in particular, it showed what it seemed to be unable to find. The other two were entirely unable to reflect that and demonstrated what I considered to be like confidence in the result when there really wasn't any.''}
Participants wished for additional clarification as to ``why'' the system performed certain actions and did not perform others [P3, P5, P10, P11]. 
P10 reflected that the highlights currently supporting this justification of system behavior \textit{``made me more annoyed at the underlying system ...  I realized there's this level of annoyance where I'm like, why didn't you do that?''}
Finally, participants attended to the sources used [P1, P2, P3, P5, P6, P9, P11, P12], raising concerns about information gleaned from AI-generated summaries on browser searches [P1, P2, P6, P9, P12] or from website titles on search pages [P2, P3, P6, P9, P11, P12].

\begin{table*}[t]
\caption{Mean rank preference for design probes and \sysname. 
Participants ranked methods by Preference (P1-P12) and by their perception of which best supported Validation (P5-P12).
For both conditions, Spec was the best and Flowchart the worst ranked, although there was much variation between participants (lower is better). The full \sysname interface was less preferred but considered more likely to support validation.
}
\label{tab:s2-pref}
\centering
\resizebox{0.35\linewidth}{!}{
\begin{tabular}{lrr}
Method      & Preference & Validation \\ \hline
Flowchart   & 2.75       & 3.25       \\
Citation    & 2.17       & 2.50       \\
Spec        & \textbf{2.08}       & 2.25       \\
\sysname & 3.00       & \textbf{2.00}      
\end{tabular}
}
\end{table*}

\subsubsection{Participants want a bird's-eye view with drill-down capabilities}
\label{sec:probes-results-birdseye}

Participants appreciated that the Flowchart [P2, P3], Citation [P12], and Spec [P4, P5] designs enabled an easy initial scan. 
P5 expressed that \textit{``I think the Spec format is more useful because it's basically summarizing the important information from the task itself.''} 
To improve scanability, P11 suggested highlighting unverified requirements and ordering assumptions by importance. 
In contrast, P5 found the Citation and P4 and P11 found the Flowchart to be challenging to scan for complex tasks.
Some participants wanted additional granularity in the Flowchart [P2, P4, P10], Citation [P10], Spec [P12], and annotations [P5]. 
Participants also wished for easier navigation to details, such as direct integration into the full trace [P3] and links to the screenshot's associated website [P6, P7, P10, P12]. 

All participants used existing attention guides like annotations, with several participants particularly expressing appreciation [P9, P10, P11, P12]. Some participants wished for more specificity with these guides, such as color matching between the Flowchart nodes and annotations [P5] and highlighting directly onto maps [P6]. 
P2, P3, and P7 both thought that searching through the trace with natural language could be helpful as well, although P1 thought that kind of interaction would add another layer of \textit{``ambiguity [and] mismatch.''}

\subsubsection{Participants varied widely in their preferences and sense of intuitiveness, with Flowchart less preferred on average}
\label{sec:probes-results-preferences}
Participants ranked the three design probes and \sysname for overall preference (P1-P12) and by which method they perceive would best help them validate (P5-P7), as reported in Table~\ref{tab:s2-pref} and Supplementary Materials. Taking the mean rank, in overall preference Spec (2.08) and Citation (2.17) were better rated than Flowchart (2.75) and \sysname (3.00). For perceived validation support, participants ranked \sysname the best (2.00) due to the level of detail, while Spec (2.25), Citation (2.50), and Flowchart (3.25) followed the same order as overall preference. 
P3 said that for \textit{``Spec and Citation, jumping back and forth between mental models was a bit of a struggle.''} P2 expressed that \textit{``I think [the Spec is] not intuitive at all and it's really more confusing... I mainly had to go check the whole trajectory myself,''} while P10 reflected that the Flowchart \textit{``feels a little unnecessary sometimes like it feels like ... I'm having to wire my brain to think in a different way.''} 
There was high variation across participants: all four methods ranked both first and last for preference and validation ability, with the exception that no participant ranked Flowchart highest for validation.
When prompted to create their ideal system, several participants mixed designs, such as Flowchart and Citation [P2], Flowchart and Spec [P5], Citation and Spec [P9], or an adaptive method dependent on the relevant task [P3, P7, P10, P11].

\subsection{Implications from the design probes study} 
\label{sec:probes-implications}

When designing a novel interface for the controlled study, we focus on the Spec condition, as it was most preferred, performed the best, includes assumptions, and enables a bird's-eye view. We want to better contextualize the Spec with the process, without creating such a focus on the process as to enable overreliance. 
To enable a high-level overview with drill-down capabilities, the new design will add additional supports to guide attention, such as linking the checklist directly to the trace rather than to disconnected screenshots and linking directly to websites. Finally, we want to differentiate between information that was not found and that which contradicts requirements.

\begin{figure*}[t]
 \centering 
 \includegraphics[width=\linewidth]{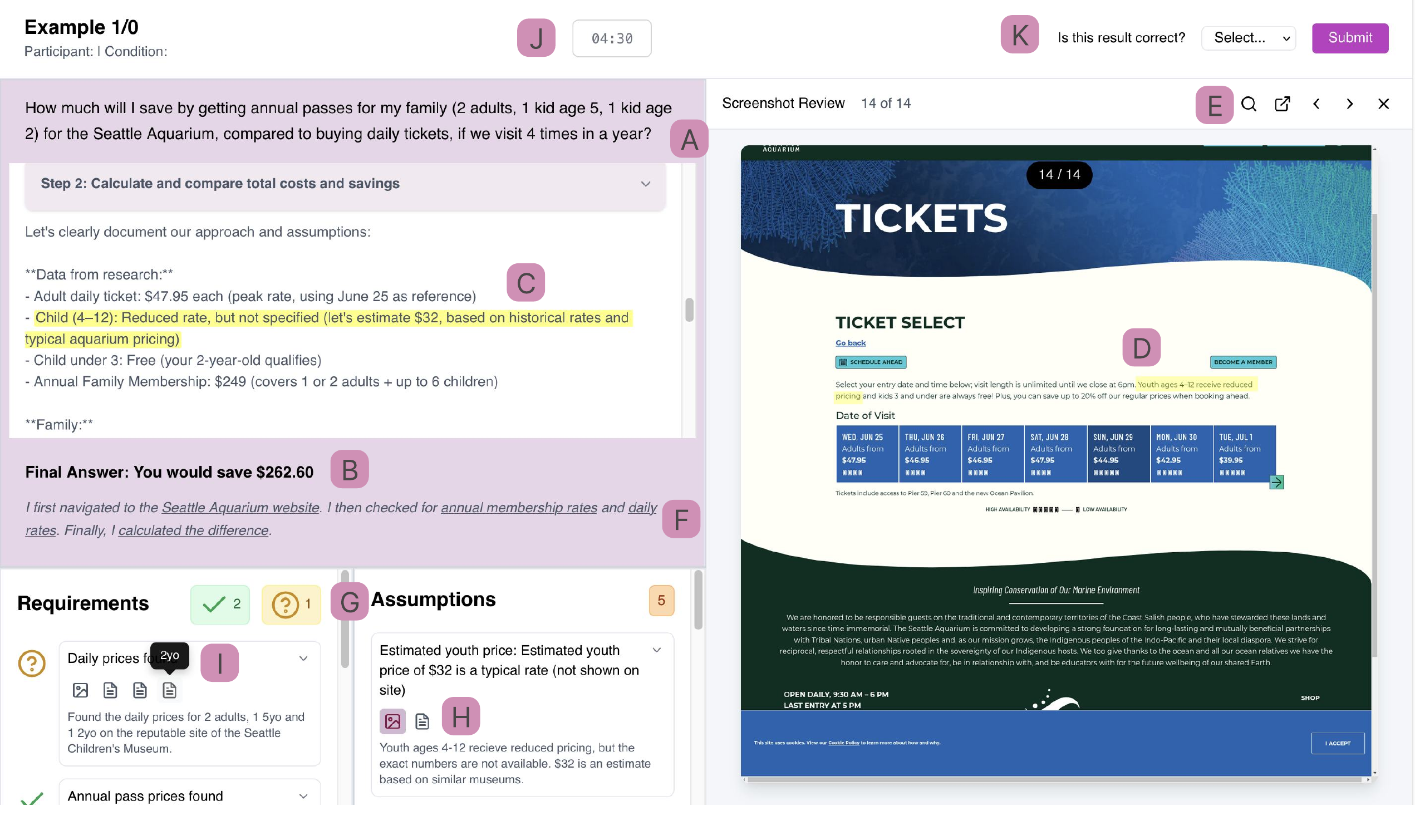}
 \caption{Controlled study interface. The interface provides the task [A] and the output to verify [B]. It integrates the \sysname trace directly into the interface, adding highlights to both the text [C] and screenshots [D]. The screenshots can be paged through, with the ability to magnify for detail and to open a new tab with the associated website for further review [E]. For additional path context, there is a brief summary [F] provided in the Citation format. For guiding review, there are requirements and assumptions [G], with additional explanatory context and icons indicating if requirements are completed, unknown, or contradicted. Each requirement and assumption refers to annotation(s) in the text and image trace with underlying icons [H] and tooltips for easier scanning [I]. For the study, there is a 5-minute timer [J] and a series of questions [K].}
 \Description{This figure shows an interface display. The top bar has a timer countdown and a question ``Is this result correct'' with a dropdown selection and a submit button. There is a pink box that contains the task text at the top left, then a white box inset that contains the trace of actions, then a final answer, then a small paragraph stating ``I first navigated to the [Seattle Aquarium website]. I then checked for [annual membership rates] and [daily rates]. Finally, I [calculated the difference]. Below this pink box, the interface is split into two columns titled ``Requirements'' and ``Assumptions. Requirements has two icons at the top, one is green with a checkmark, and the number 2, and the other is yellow with a question mark and the number 1. Next to assumptions is an orange number 5. Below each are boxes containing a requirement or assumption. Each box has the text (e.g., ``Estimated youth price of \$32 is a typical rate (not shown on site)'') and several icons below, formatted either as an image icon or a document icon. Hovered icons show a black tooltip, in this case ``2yo''. Selected icons turn a magenta color. Below the icons is a brief explanatory description (e.g., ``Youth ages 4-12 receive reduced pricing, but the exact numbers are not available. \$32 is an estimate based on similar museums.'' The right side has a header titled ``Screenshot Review'' indicating image 14 of 14. This header has a magnification icon, a browser icon, and arrow keys. The image itself has yellow highlighting. The text in the trace also has yellow highlighting. }
\label{fig:combinedinterface} 
\end{figure*}

\section{Controlled study}
\label{sec:control}

\yellowbox{Takeaway: We implement a specification that guides navigation to detailed steps and surfaces assumptions. This approach reduced the time for error finding but increased false confidence.}

We next create an interface combining useful elements from the design probes and adding additional linking (\ref{sec:control-system}). We test this interface with \sysname in a counterbalanced, Wizard-of-Oz study (\ref{sec:control-procedure}) and measure its effect on accuracy, speed, and user experience (\ref{sec:control-results}).

\subsection{System design}
\label{sec:control-system}

The baseline condition recreates the Magentic-UI interface for similarity across conditions. All screenshots, all agent text, the final answer summary, and the icons by text steps navigating to a screenshot are visible and functional. However, there are no longer multiple traces, a chat function, the live browser, or a connection to agentic functions. The only added function is a magnification glass icon that enables a visual inset magnifying the screenshot. 

The treatment condition contains the \sysname trace with several updates (Figure~\ref{fig:combinedinterface}). There is a magnifying glass icon and an icon linking to the relevant website [E] as well as highlighted annotations on the images [D] and the text [C]. Above the text trace is the task [A], and below is the output [B]. There is also a short summary of the path taken in the Citation format [F].
Further below are requirements and assumptions, each with a brief explanation [G]. There are also icons indicating image(s), text snippet(s), and code snippet(s) that can be clicked, relating to that requirement or assumption [H]. Each icon has a tool tip with a brief description [I]. Requirements have a green check mark, a yellow question mark, or a red X to indicate if the requirement was completed, is unknown, or is contradicted, respectively. At the top of each requirement and assumption section, there is a count of the number of requirements of each type and of the number of assumptions. 

The treatment condition also provides linking among interface elements. Users can select specification icons, the highlighted text, or the image annotations. Selecting one element highlights and scrolls to the relevant screenshot, text trace, and specification icon. Since there can be many annotations per screenshot, only relevant annotations are shown during a selection.
Both conditions have a horizontal header at the top of the screen that contains a five-minute timer [J] and the questions for the participants to answer [K].

\subsection{Procedure}
\label{sec:control-procedure}

This study involved 12 participants, all employees of a large tech firm with some experience with agentic systems. No participants overlapped with the first two studies. Ten participants were in person, and two were virtual. All interacted with a pre-installed system. 
Participants first signed a consent form and reviewed a demonstration of \sysname's function as a CUA. We instructed participants to imagine they had input these tasks, were returning after the agent completed, and needed to decide if they thought the output was correct. 

We had two conditions: baseline and treatment. Each participant saw both conditions and all tasks, with the order of these conditions and their associated tasks balanced across all participants. In each set, two outputs were correct, and two were incorrect. As in the previous study, the tasks were selected from AssistantBench~\cite{yoran2024assistantbench}, prioritizing those that cannot be validated by a single screenshot. They are available in Supplementary Materials. We used actual runs of \sysname using OpenAI's GPT-5~\cite{openai2025gpt5}, a state-of-the-art model at the time of the study. The annotations were created by hand based on \sysname's reasoning, such that if the trace incorrectly states that a requirement was completed, the interface will say that the requirement was completed. 

For each task, participants see the task, answer, and interface. They are allowed to use the browser with no restrictions. We kept a soft deadline of 5 minutes per task in order to ensure all tasks were completed within the study time limits. At the top of the display, there was a timer that turned red after 5 minutes. If participants did not notice, the study conductor gave them a gentle reminder. 
Participants then answered ``Yes'' or ``No'' to the question ``Is this result correct?'' They then selected confidence on a 1-7 scale from ``Not at all confident'' to ``Extremely confident'' in response to the question ``How confident are you in your previous answer?'' Finally, they gave a free-form answer describing ``Why or why not?''
We did not require participants to think aloud. At the conclusion of the study, we perform a semi-structured interview for the remaining time of approximately 10 minutes. Studies were one hour long. 

We measure if participants correctly determine if the output was correct, their subjective confidence score, and how long it takes for them to answer. The duration measurement ends after participants submit their result correctness guess, not including the time to input their confidence scores and justification. Qualitatively, we review their justification responses, any think-aloud comments that were made, and responses to the semi-structured interview. We use line-by-line coding by hand, then perform another two rounds of inductive and deductive thematic analysis~\cite{braun2006using}.

In line with recommendations for fair statistical communication in HCI~\cite{dragicevic2016fair}, especially as our study does not have high power due to resource constraints, we report effect sizes and confidence intervals as they show magnitude and do not rely on sample size (Table~\ref{tab:s3adc}). 
As the study is counterbalanced, we calculate Cohen's $d_{av}$, a paired measurement. We additionally report results for subsets of data in which the ground truth is correct or incorrect and user responses are correct or incorrect. In these subsets, sample sizes are unequal, so we use Cohen's $d_{ind}$. 
We apply a correction factor to Cohen's $d$ to use the unbiased Hedges' $g$, and we use cutoffs of 0.2, 0.5, and 0.8 for small, medium, and large effect sizes, respectively. We calculate the 95\% confidence interval using the noncentral $t$ distribution. We use calculations as specified in prior work~\cite{goulet2018review}.

\begin{figure*}[t]
 \centering 
 \includegraphics[width=\linewidth]{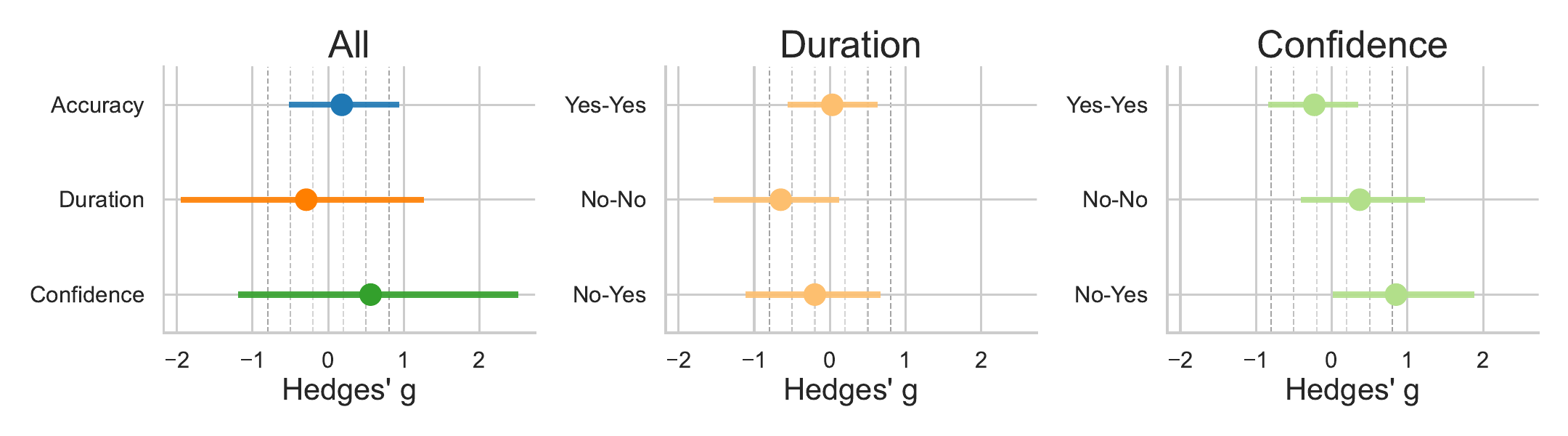}
 \caption{Control study results. The chart shows the Hedges' $g$ effect size and the 95\% confidence interval. The Duration and Confidence charts show subsets such that ``No-Yes'' displays when the ground truth answer is ``No'' and the user answers ``Yes''. 
 Overall, we find no effect on accuracy (0.18), a small effect decreasing duration (-0.29), and a medium effect increasing confidence (0.56). The effect on duration is medium-sized when the user correctly finds an error (``No-No'': -0.65). The largest confidence effect occurs when the answer is incorrect, but the user thinks it is correct (``No-Yes'': 0.85). Exact values are in Table~\ref{tab:s3adc}. Three dotted vertical lines show the cutoffs for small, medium, and large effect sizes.}
 \Description{Three graphs. Values are from the main results table. Each has Hedges' g on the x-axis. First graph titled ``All'' has y-axis values ``Accuracy'' ``Duration'' and ``Confidence''. Second graph titled ``Duration'' and third graph titled ``Confidence'' have y axis values ``Yes-Yes'', ``No-No'', and ``No-Yes''. Each graph has three points indicating the effect size with horizontal bars spanning the 95\% confidence interval. There are light grey bars for small, medium, and large effect sizes at 0.2, 0.5, and 0.8.}
\label{fig:effectsizes} 
\end{figure*}

\subsection{Results}
\label{sec:control-results}
We report the impact of the treatment condition on the user experience and duration (\ref{sec:control-results-duration}) and on agreement and confidence (\ref{sec:control-results-agreement}). We then discuss use of the interface elements (\ref{sec:control-results-r/a/linking}) and opportunities for improved detail (\ref{sec:control-results-moredetail}).

\subsubsection{User experience and time demands improve, with room for growth}
\label{sec:control-results-duration}

We find a small effect size of duration (Hedges' $g$: -0.29) in which the treatment condition is faster. As seen in Figure~\ref{fig:effectsizes}, the bulk of this difference occurs for questions in which the user correctly answers the question (Hedges' $g$: -0.65). 
Participants appreciated that the treatment condition had more support in navigating large amounts of information than the baseline [P1, P2, P6, P9, P10, P12].
Participants especially appreciated the connection to the browser site [P3, P7, P12]. 
P6 shared their experience in the baseline condition: \textit{``For me to understand what assumptions it has done, I have to read through a lot. I appreciated in the [treatment] UI that it was clearly shown. For instance, [there were] options for choosing other breeds for the dog, but it chose the most common.''} 
Despite the improvements, participants still found that the amount to verify was overwhelming [P1, P3, P5, P9, P11] with much text [P2, P9, P10] and a high learning curve to understand all the components [P1, P4, P5, P7, P9, P10]. 
To improve this overwhelming information, participants suggested color-coded highlighting [P6, P9], condensing the number of information panes [P1, P5], and the ability to directly ask questions about justification to the CUA through a chat interface [P1, P2, P10].

\subsubsection{Our interface did not meaningfully improve agreement but did increase confidence when participants were incorrect.} 
\label{sec:control-results-agreement}

We find that the treatment condition does not substantially improve participants' discretion of whether the output is correct or incorrect (Hedges' $g$: 0.18). 
However, we find a medium effect size (Hedges' $g$: 0.56) that users are more confident in their responses in the treatment condition. The effect is largest when the output is incorrect, but participants score it as correct (Hedges' $g$: 0.85). 
Participants also mentioned a lack of trust in themselves to verify the outputs in day-to-day life. P5 noticed that they \textit{``can shut [their] brain off when using these systems.''}
P3 predicted that since there are \textit{``too many requirements, I would check some, but [I] still think it's good it's provided to me.''} 
Participants often checked portions of the task in which correctness would be clear, such as math calculations [P2, P3, P4, P6, P7, P8, P10, P11, P12] or if the screenshots matched the text [P1, P2, P7, P10], while missing errors in another part of the process.

\subsubsection{Participants used and appreciated the requirements, assumptions, and linking}
\label{sec:control-results-r/a/linking}

Many participants reflected that they relied heavily on requirements when verifying [P1, P2, P3, P5, P6, P9]. 
P6 shared \textit{``when I think about like the requirements here, this is like the natural language way of formulating how you would like do a search [like] you would in a database, for example. And I find it really helpful to let me reason through whether this makes sense or not.''} 
Participants also utilized the assumptions, although they noted that assumptions would be more useful in day-to-day applications than they were in the study [P1, P2, P4, P5, P9, P12]. 
Participants justified their decision with plan reasonability both when they were correct [P3, P5, P10, P9] and incorrect [P5, P9]. 
P9 also explicitly mentioned the reasonableness of requirements and assumptions as motivation: \textit{``all the planning steps make sense to me. The requirements, they make sense to me. The assumptions make sense to me as well. So I'm going to go ahead. And I feel the results are correct.''}

All participants used the annotations and tooltips, but they were especially noted as helpful by P10, who struggled to find a small piece of information on a busy screenshot during the baseline condition.
Similarly, P4 mentioned \textit{``I do like that the annotation pointed it out to me, though, because I might have missed it.''} 
In contrast, most participants did not use the short citation-style summary of the workflow.
Participants also appreciated the linking between different parts of the UI [P5, P10] and to the websites themselves [P2, P3, P7, P12] that assisted navigation. 
All participants used the text trace at some point, in contrast to only half of the participants in the previous study. 
However, P6 found tracking changes across all parts of the interface difficult, and P5 disliked that selecting an item hid other annotations, as they would forget to deselect a selected item.

\subsubsection{The specification can improve with more detail and indications of global completeness.}
\label{sec:control-results-moredetail}

Some participants thought the existing requirements and assumptions were sufficient [P3, P5, P6, P9]. 
Others requested additional granularity in the given requirements [P2, P6, P9], assumptions [P1, P4, P12], and screenshots [P2, P11].
Additionally, several participants thought they would still need access to a search engine to sanity check and confirm if information is missing [P3, P4, P10, P12]. 
P5 expressed a lack of trust in the system's all-or-nothing self-declaration of complete requirements: \textit{``I don't know if I would trust the confidence, but it doesn't really have a confidence. It just seems like it's 100\% confident [with] the check mark.''}

Participants also requested the ability to modify requirements and assumptions and have that impact task execution [P2, P5, P9]. 
P1, P6, and P9 disagreed with the assumptions given so would have changed them if completing the task for their own standards. Relatedly, participants raised several suggestions for improving assumptions. 
When there were minimal or no assumptions, P1 and P12 raised concern that the overview was likely missing details. 
P4 and P2 considered how assumption content could be task dependent (e.g., gathering information, reasoning, or taking actions). 
P4 wondered how assumptions could incorporate priors in a broader context.

All participants expressed on at least one task that the CUA was not sufficiently thorough. 
P4, P11, and P12 wondered if there were more legitimate sources that were missing from \sysname's investigation. 
P5 reflected \textit{``I'm like sort of trusting the model is doing this complete search, but I can't really check that. I can only check like certain facts if they're right or wrong.''}

\begin{table*}[t]
\caption{Quantitative results between the Baseline and Treatment conditions. GT and User refer to the ground truth and user answer to the question      ``Is this output correct?'' respectively. ``All'' includes both ``Yes'' and ``No'' responses. Effects compare the Baseline (Base.) and Treatment (Treat.) conditions. We report the type of Cohen's $d$ calculated (Cohen's) and all effect sizes (Eff.) have an unbiasing correction applied to become Hedges' $g$. We include the lower (CI L) and upper (CI U) confidence intervals. 
Dashes occur where the effect cannot be calculated.
Confidence is measured on a 7-point Likert scale. }
\label{tab:s3adc}
\centering
\resizebox{\linewidth}{!}{
\begin{tabular}{llllrrrrrrrrrrrrrrr}
                     &                      &                         &          & \multicolumn{5}{l}{Accuracy (\%)}                                                       & \multicolumn{5}{l}{Duration (s)}                                                         & \multicolumn{5}{l}{Confidence (1-7)}                                                  \\
GT                   & User                 & Cohen's                 & Cond.    & Mean   & STD   & Eff.                  & CI L                   & CI H                  & Mean   & STD   & Eff.                 & CI L                   & CI H                  & Mean & STD  & Eff.                 & CI L                   & CI H                  \\ \hline
\multirow{2}{*}{No}  & \multirow{2}{*}{No}  & \multirow{2}{*}{$d_{ind}$} & Base.    & 100.00 & 0.00  & --                    & --                     & --                    & 270.79 & 62.93 & \multirow{2}{*}{-0.65} & \multirow{2}{*}{-1.54} & \multirow{2}{*}{0.12} & 3.54 & 1.85 & \multirow{2}{*}{0.37}  & \multirow{2}{*}{-0.41} & \multirow{2}{*}{1.23} \\
                     &                      &                         & Treat.   & 100.00 & 0.00  & --                    & --                     & --                    & 229.63 & 60.49 &                        &                        &                       & 4.23 & 1.74 &                        &                        &                       \\ \hline
\multirow{2}{*}{No}  & \multirow{2}{*}{Yes} & \multirow{2}{*}{$d_{ind}$} & Base.    & 0.00   & 0.00  & --                    & --                     & --                    & 251.85 & 93.01 & \multirow{2}{*}{-0.20} & \multirow{2}{*}{-1.12} & \multirow{2}{*}{0.67} & 3.64 & 1.63 & \multirow{2}{*}{0.85}  & \multirow{2}{*}{0.01}  & \multirow{2}{*}{1.88} \\
                     &                      &                         & Treat.   & 0.00   & 0.00  & --                    & --                     & --                    & 234.48 & 69.99 &                        &                        &                       & 4.82 & 0.98 &                        &                        &                       \\ \hline
Yes                  & No                   & $d_{ind}$                  & Base. & 0.00   & 0.00  & --                    & --                     & --                    & 151.31 & 98.17 & --                     & --                     & --                    & 1.50 & 0.71 & --                     & --                     & --                    \\ \hline
\multirow{2}{*}{Yes} & \multirow{2}{*}{Yes} & \multirow{2}{*}{$d_{ind}$} & Base.    & 100.00 & 0.00  & --                    & --                     & --                    & 208.39 & 61.40 & \multirow{2}{*}{0.03}  & \multirow{2}{*}{-0.56} & \multirow{2}{*}{0.63} & 5.05 & 1.17 & \multirow{2}{*}{-0.23} & \multirow{2}{*}{-0.84} & \multirow{2}{*}{0.35} \\
                     &                      &                         & Treat.   & 100.00 & 0.00  & --                    & --                     & --                    & 210.71 & 86.15 &                        &                        &                       & 4.71 & 1.63 &                        &                        &                       \\ \hline
\multirow{2}{*}{All} & \multirow{2}{*}{All} & \multirow{2}{*}{$d_{av}$}  & Base.    & 72.92  & 44.91 & \multirow{2}{*}{0.18} & \multirow{2}{*}{-0.52} & \multirow{2}{*}{0.94} & 232.87 & 76.07 & \multirow{2}{*}{-0.29} & \multirow{2}{*}{-1.95} & \multirow{2}{*}{1.26} & 4.17 & 1.71 & \multirow{2}{*}{0.56}  & \multirow{2}{*}{-1.19} & \multirow{2}{*}{2.51} \\
                     &                      &                         & Treat.   & 77.08  & 42.47 &                       &                        &                       & 221.28 & 75.67 &                        &                        &                       & 4.60 & 1.53 &                        &                        &                      
\end{tabular}
}
\end{table*}
\section{Discussion}
\label{sec:discussion}

In three user studies, we find that people miss errors with the existing trace display of listing and summarizing execution steps. We then explore the design space for improved trace design and elucidate challenges in human verification of multi-agent workflows.
In this section, we discuss our findings and propose future work regarding trace construction (\ref{sec:discussion-verbose}), focus on the agent's process (\ref{sec:discussion-process}), extensions beyond the agent's list of actions (\ref{sec:discussion-didnotdo}), participants' definitions of correctness (\ref{sec:discussion-correctness}), and opportunities beyond post-hoc verification (\ref{sec:discussion-running}).

\subsection{Error finding requires detail, but lists of actions are overwhelmingly verbose. Supports should provide navigation for this detail to instill appropriate confidence.} 
\label{sec:discussion-verbose}

We find that participants miss errors and feel overwhelmed when reviewing long traces of an agent's series of actions. 
Summarizing sets of steps reduces this verbosity but can hide execution errors. 
Thus, determining the correct level of granularity for reporting agent actions is a crucial challenge~\cite{kambhamettu2025attribution,shavit2023practices,feng2025regulatory,shome2025johnny,bansal2024challenges}. 
Rather than adjusting the granularity of the step detail, our approach supports user navigation of execution detail with a higher-level specification to reduce user effort while staying grounded in execution steps.  
However, there remain many open questions towards supporting this navigation. 
Our treatment interface had a high learning curve and increased confidence, especially when users were wrong.

Future investigations can study verification practices with more ecological validity to everyday use of these agents.
In our controlled study, participants had access to the browser, so many executed the task themselves. This ability to complete the task themselves may have evened performance between the two conditions.
However, several participants desired lightweight validation because they would not extend such effort in everyday life, especially for low-stakes tasks. 

Future work can also investigate if this higher confidence translates to worse decision-making and study the integration of methods to reduce overreliance, such as cognitive forcing functions~\cite{buccinca2021trust}.
We included checked requirements as a part of the design to improve scanability. 
However, incorrectly confirmed requirements might contribute to higher confidence as they are a highly visible set of green checkmarks. 
Further iteration on requirement and assumption content and implementation could improve existing outcomes. 
Finally, the space of opportunities for trace navigation depends on the system's ability to generate these requirement,  assumption, and annotation guides.

\subsection{Process provenance matters but is challenging to effectively display.}
\label{sec:discussion-process}

Across all studies, participants relied on the overall process and cared about the quality of information sources.
However, a high-level review of the process could hide details in the execution of individual steps. 
In the design probes study, participants often referred to a ``reasonable'' process when justifying why they thought an incorrect answer was correct. 
Despite this risk of overreliance, many participants found interpreting the results without the process's global context to be challenging. 
The Specification method had the highest accuracy, but it took the longest, required the most checks of the overall trace, and some participants lamented its disconnect from the process.

To navigate this tension, our final design's main feature was the outcome-oriented Specification approach linked directly into the trace. 
Participants justified overreliance with a reasonable process less often in the controlled study than in the design probes study.
Future work could investigate if the challenge of being divorced from the process might positively encourage more cognitive engagement and run additional control studies on content inclusion for different presentation methods, such as integrating assumptions into a flowchart design.

\subsection{Trace displays can extend to show what the agent did not do.}
\label{sec:discussion-didnotdo}

In the open world of browser use, there can be many approaches to answering a given task, especially when the task is underspecified. 
Participants were aware that the CUA was choosing one path among many and wanted this broader context to be more explicitly surfaced. 
We surfaced this information by listing the assumptions that implicitly raise alternative paths and by distinguishing between requirements that were contradicted and those that did not have sufficient information found. 
Additional context could include highlighting references, summarizing dead-end paths, and testing multiple paths for a range of outputs.

Many assumptions made by the agent occur at decision points that cannot be predicted during the co-planning stage. Clarifying questions during runtime can help with steering and received positive feedback from participants. However, we expect that asking a clarifying question for all decisions made, such as whether a certain source is trustworthy, would be overburdening. Future work can further develop thresholds for what decisions warrant interrupting the user and methods to make the user effectively aware of assumptions that did not meet the threshold. 

\subsection{Participants display subjective, changing, and personal definitions of correctness.}
\label{sec:discussion-correctness}

We observe that people change their minds partway through execution, have subjective views of what quality counts as ``good enough,'' and like to explore multiple paths. Additionally, people can hold personal context and preferences they may not think to disclose. 
In contrast, most benchmarks measuring agentic performance assume one correct answer~\cite{yoran2024assistantbench,mialon2023gaia,deng2023mind2web}. In the examples we used in which multiple answers were reasonable, the benchmark's answer defaulted to the easiest path. 
For instance, for a task about comparing shipping prices, the benchmark assumes that the user is not shipping to a residential address, as changing that assumption would require checking a tick box~\cite{yoran2024assistantbench}. For the Seattle Children's Museum, which displays all prices on one website page, \sysname's path consistently assumed that the user was a Washington resident. However, for the Seattle Aquarium, which requires checking a tick box to claim residency, \sysname's path consistently assumed that the user was not a Washington resident.
Benchmarks with simulated users incorporate some degree of this flexibility, such as by dynamically updating the answer within a restricted set of domain policies~\cite{yao2024tau} or by having an LLM judge correctness on a 1-5 scale~\cite{shao2024collaborative}.

Measuring performance and comparing systems is much easier when limited to the subset of tasks and assumptions that can be easily deemed correct or incorrect. However, this binary construction does not fully represent how users actually interact with agents. We see rich opportunity for evaluation methods that explicitly measure adaptability to changing or underspecified user goals, especially as the CUA struggled to backtrack and adapt to user requests.

\subsection{Verification mechanisms could integrate into a running agentic workflow.}
\label{sec:discussion-running}

As our study focuses on measuring error-finding capabilities, we assume a multi-tasking scenario in which the user returns to a completed query for post-hoc verification. 
There are abundant opportunities to extend this work to semi-autonomous verification during workflow execution. 
For instance, participants requested the ability to add, reject, and change requirements during the co-planning stage and to change assumptions retroactively. Such interaction could also include the specification of both hard and soft requirements that an automated validation agent then tracks as the workflow runs. 
The ability to chat with the trace would enable specific querying of the source of different statements.
Finally, after finding an error, future work can enable changing the outcome without re-running the entire chain. Such an interaction paradigm could enable users to return to a previously completed step, make a change, and continue execution from that checkpoint. 

\section{Limitations}

Our study has several limitations beyond those discussed in the previous section.
Our participants were all employees of a large technology firm who had previously used AI agents. 
These participants are not representative of the broader set of users with access to publicly released agents.
Additionally, we only test one CUA system, \sysname. 
Other reasoning models and agents outside of CUA similarly use long traces as explanation, so extending a similar design process to those will improve understanding of these findings' generalizability.
Our study setup also reviewed just one part of verification, post-hoc verification with an assumed correct or incorrect answer. 
Finally, our metric of confidence was a self-defined scale from 1 to 7. 
Further exploring this strong effect with a more nuanced metric of confidence would better define this issue.

We ran a Wizard-of-Oz study to enable rapid iteration on the system design. However, full implementation of this interface would require automated surfacing of the abstractions. 
After predicting requirements as a part of the planning stage, automated judges could track if requirements are completed and what assumptions occur, similar to how \sysname's action guard acts as an external evaluator. Highlighted annotations would also require an external agent for post-hoc determination of relevant information. Future work may consider alternatives if information is unavailable on the screenshot. 
Existing systems that currently display a list of actions could then also display an always visible, outcome-oriented overview that updates as the agent makes progress. 
A fully implemented system may make mistakes and output different annotations than a WoZ implementation, so its efficacy in supporting human oversight must also be evaluated. 

\section{Conclusion}

Through a formative study, we found that people miss small but impactful errors in the verbose existing CUA system and have a subjective and changing sense of correctness. 
A follow-up study of three design probes revealed that people could overrely if the overall process is reasonable and that they wanted a high-level overview with drill-down capabilities.  
From these findings, we create a novel interface and test its impact on participants' correctness judgments in a counterbalanced study.  
With our interface, users found errors faster, but had higher confidence without meaningfully improving accuracy. 
Our studies illuminate challenges and opportunities for future development supporting human oversight of multi-agent systems.

\begin{acks}
Thank you to the Microsoft Research AI Frontiers group for their feedback.
\end{acks}

\bibliographystyle{ACM-Reference-Format}
\bibliography{0_main}

@String{Computing = "Computing" }

@String{Computer = "{IEEE} Computer" }

@String{Springer = "Springer-Verlag" }

@ArtifactSoftware{R,
    title = {R: A Language and Environment for Statistical Computing},
    author = {{R Core Team}},
    organization = {R Foundation for Statistical Computing},
    address = {Vienna, Austria},
    year = {2019},
    url = {https://www.R-project.org/},
}

@inbook{brooke1996sus,
author = {Brooke, John},
year = {1996},
month = {01},
pages = {189-194},
title = {SUS -- a quick and dirty usability scale}
}

@article{shome2025johnny,
  title={Why Johnny Can't Use Agents: Industry Aspirations vs. User Realities with AI Agent Software},
  author={Shome, Pradyumna and Krishnan, Sashreek and Das, Sauvik},
  journal={arXiv preprint arXiv:2509.14528},
  year={2025}
}

@article{feng2024cocoa,
  title={Cocoa: Co-planning and co-execution with ai agents},
  author={Feng, KJ and Pu, Kevin and Latzke, Matt and August, Tal and Siangliulue, Pao and Bragg, Jonathan and Weld, Daniel S and Zhang, Amy X and Chang, Joseph Chee},
  journal={arXiv preprint arXiv:2412.10999},
  year={2024}
}

@article{shao2024collaborative,
  title={Collaborative gym: A framework for enabling and evaluating human-agent collaboration},
  author={Shao, Yijia and Samuel, Vinay and Jiang, Yucheng and Yang, John and Yang, Diyi},
  journal={arXiv preprint arXiv:2412.15701},
  year={2024}
}

@inproceedings{huq2025cowpilot,
    title = "{C}ow{P}ilot: A Framework for Autonomous and Human-Agent Collaborative Web Navigation",
    author = "Huq, Faria  and
      Wang, Zora Zhiruo  and
      Xu, Frank F.  and
      Ou, Tianyue  and
      Zhou, Shuyan  and
      Bigham, Jeffrey P.  and
      Neubig, Graham",
    editor = "Dziri, Nouha  and
      Ren, Sean (Xiang)  and
      Diao, Shizhe",
    booktitle = "Proceedings of the 2025 Conference of the Nations of the Americas Chapter of the Association for Computational Linguistics: Human Language Technologies (System Demonstrations)",
    month = apr,
    year = "2025",
    address = "Albuquerque, New Mexico",
    publisher = "Association for Computational Linguistics",
    url = "https://aclanthology.org/2025.naacl-demo.17/",
    doi = "10.18653/v1/2025.naacl-demo.17",
    pages = "163--172",
    ISBN = "979-8-89176-191-9"
}

@misc{openai2025operator,
  author       = {OpenAI},
  title        = {Introducing Operator},
  year         = {2025},
  howpublished = {\url{https://openai.com/index/introducing-operator/}}
}

@article{awadallah2025fara,
  title={Fara-7B: An Efficient Agentic Model for Computer Use},
  author={Awadallah, Ahmed and Lara, Yash and Magazine, Raghav and Mozannar, Hussein and Nambi, Akshay and Pandya, Yash and Rajeswaran, Aravind and Rosset, Corby and Taymanov, Alexey and Vineet, Vibhav and others},
  journal={arXiv preprint arXiv:2511.19663},
  year={2025}
}

@misc{openai2024gpt4o,
  author       = {OpenAI},
  title        = {Hello GPT-4o},
  year         = {2024},
  howpublished = {\url{https://openai.com/index/hello-gpt-4o/}}
}

@misc{openai2025gpt5,
  author       = {OpenAI},
  title        = {Introducing GPT-5},
  year         = {2025},
  howpublished = {\url{https://openai.com/index/introducing-gpt-5/}}
}

@article{collins2024building,
  title={Building machines that learn and think with people},
  author={Collins, Katherine M and Sucholutsky, Ilia and Bhatt, Umang and Chandra, Kartik and Wong, Lionel and Lee, Mina and Zhang, Cedegao E and Zhi-Xuan, Tan and Ho, Mark and Mansinghka, Vikash and others},
  journal={Nature human behaviour},
  volume={8},
  number={10},
  pages={1851--1863},
  year={2024},
  publisher={Nature Publishing Group UK London}
}

@article{mozannar2025magentic,
  title={Magentic-ui: Towards human-in-the-loop agentic systems},
  author={Mozannar, Hussein and Bansal, Gagan and Tan, Cheng and Fourney, Adam and Dibia, Victor and Chen, Jingya and Gerrits, Jack and Payne, Tyler and Maldaner, Matheus Kunzler and Grunde-McLaughlin, Madeleine and others},
  journal={arXiv preprint arXiv:2507.22358},
  year={2025}
}

@article{kapoor2024ai,
  title={Ai agents that matter},
  author={Kapoor, Sayash and Stroebl, Benedikt and Siegel, Zachary S and Nadgir, Nitya and Narayanan, Arvind},
  journal={arXiv preprint arXiv:2407.01502},
  year={2024}
}

@inproceedings{bennett2023does,
  title={How does HCI understand human agency and autonomy?},
  author={Bennett, Dan and Metatla, Oussama and Roudaut, Anne and Mekler, Elisa D},
  booktitle={Proceedings of the 2023 CHI Conference on Human Factors in Computing Systems},
  pages={1--18},
  year={2023}
}

@inproceedings{chan2023harms,
  title={Harms from increasingly agentic algorithmic systems},
  author={Chan, Alan and Salganik, Rebecca and Markelius, Alva and Pang, Chris and Rajkumar, Nitarshan and Krasheninnikov, Dmitrii and Langosco, Lauro and He, Zhonghao and Duan, Yawen and Carroll, Micah and others},
  booktitle={Proceedings of the 2023 ACM Conference on Fairness, Accountability, and Transparency},
  pages={651--666},
  year={2023}
}

@article{kulveit2025gradual,
  title={Gradual disempowerment: Systemic existential risks from incremental AI development},
  author={Kulveit, Jan and Douglas, Raymond and Ammann, Nora and Turan, Deger and Krueger, David and Duvenaud, David},
  journal={arXiv preprint arXiv:2501.16946},
  year={2025}
}

@article{feng2025levels,
  title={Levels of Autonomy for AI Agents},
  author={Feng, KJ and McDonald, David W and Zhang, Amy X},
  journal={arXiv preprint arXiv:2506.12469},
  year={2025}
}

@inproceedings{feng2025regulatory,
  title={On the Regulatory Potential of User Interfaces for AI Agent Governance},
  author={Feng, Kevin and Kim, Tae Soo and Pang, Rock Yuren and Huq, Faria and August, Tal and Zhang, Amy X},
  booktitle={NeurIPS 2025 Workshop on Regulatable ML}
}

@article{shavit2023practices,
  title={Practices for governing agentic AI systems},
  author={Shavit, Yonadav and Agarwal, Sandhini and Brundage, Miles and Adler, Steven and O’Keefe, Cullen and Campbell, Rosie and Lee, Teddy and Mishkin, Pamela and Eloundou, Tyna and Hickey, Alan and others},
  journal={Research Paper, OpenAI},
  year={2023}
}

@inproceedings{balepur2025good,
  title={A Good Plan is Hard to Find: Aligning Models with Preferences is Misaligned with What Helps Users},
  author={Balepur, Nishant and Shu, Matthew and Sung, Yoo Yeon and Goldfarb-Tarrant, Seraphina and Feng, Shi and Yang, Fumeng and Rudinger, Rachel and Boyd-Graber, Jordan Lee},
  booktitle={Proceedings of the 2025 Conference on Empirical Methods in Natural Language Processing},
  pages={11579--11606},
  year={2025}
}

@article{ibrahim2025measuring,
  title={Measuring and mitigating overreliance is necessary for building human-compatible AI},
  author={Ibrahim, Lujain and Collins, Katherine M and Kim, Sunnie SY and Reuel, Anka and Lamparth, Max and Feng, Kevin and Ahmad, Lama and Soni, Prajna and Kattan, Alia El and Stein, Merlin and others},
  journal={arXiv preprint arXiv:2509.08010},
  year={2025}
}

@inproceedings{song2025human,
  title={Human-AI Collaboration with Misaligned Preferences},
  author={Song, Jiaxin and Shahkar, Parnian and Donahue, Kate and Chaudhury, Bhaskar Ray},
  booktitle={Proceedings of the 5th ACM Conference on Equity and Access in Algorithms, Mechanisms, and Optimization},
  pages={294--294},
  year={2025}
}

@article{kambhamettu2025attribution,
  title={Attribution Gradients: Incrementally Unfolding Citations for Critical Examination of Attributed AI Answers},
  author={Kambhamettu, Hita and Hwang, Alyssa and Laban, Philippe and Head, Andrew},
  journal={arXiv preprint arXiv:2510.00361},
  year={2025}
}

@article{zou2025survey,
  title={A survey on large language model based human-agent systems},
  author={Zou, Henry Peng and Huang, Wei-Chieh and Wu, Yaozu and Chen, Yankai and Miao, Chunyu and Nguyen, Hoang and Zhou, Yue and Zhang, Weizhi and Fang, Liancheng and He, Langzhou and others},
  journal={Authorea Preprints},
  year={2025},
  publisher={Authorea}
}

@article{mitchell2025fully,
  title={Fully autonomous ai agents should not be developed},
  author={Mitchell, Margaret and Ghosh, Avijit and Luccioni, Alexandra Sasha and Pistilli, Giada},
  journal={arXiv preprint arXiv:2502.02649},
  year={2025}
}

@article{zou2025call,
  title={A Call for Collaborative Intelligence: Why Human-Agent Systems Should Precede AI Autonomy},
  author={Zou, Henry Peng and Huang, Wei-Chieh and Wu, Yaozu and Miao, Chunyu and Li, Dongyuan and Liu, Aiwei and Zhou, Yue and Chen, Yankai and Zhang, Weizhi and Li, Yangning and others},
  journal={arXiv preprint arXiv:2506.09420},
  year={2025}
}

@article{chang2025chatbench,
  title={Chatbench: From static benchmarks to human-ai evaluation},
  author={Chang, Serina and Anderson, Ashton and Hofman, Jake M},
  journal={arXiv preprint arXiv:2504.07114},
  year={2025}
}

@inproceedings{maulsby1993prototyping,
  title={Prototyping an intelligent agent through Wizard of Oz},
  author={Maulsby, David and Greenberg, Saul and Mander, Richard},
  booktitle={Proceedings of the INTERACT'93 and CHI'93 conference on Human factors in computing systems},
  pages={277--284},
  year={1993}
}

@article{shoham1993agent,
  title={Agent-oriented programming},
  author={Shoham, Yoav},
  journal={Artificial intelligence},
  volume={60},
  number={1},
  pages={51--92},
  year={1993},
  publisher={Elsevier}
}

@inproceedings{lieberman1997autonomous,
author = {Lieberman, Henry},
title = {Autonomous interface agents},
year = {1997},
isbn = {0897918029},
publisher = {Association for Computing Machinery},
address = {New York, NY, USA},
url = {https://doi.org/10.1145/258549.258592},
doi = {10.1145/258549.258592},
booktitle = {Proceedings of the ACM SIGCHI Conference on Human Factors in Computing Systems},
pages = {67–74},
numpages = {8},
location = {Atlanta, Georgia, USA},
series = {CHI '97}
}

@article{maes1994agents,
author = {Maes, Pattie},
title = {Agents that reduce work and information overload},
year = {1994},
issue_date = {July 1994},
publisher = {Association for Computing Machinery},
address = {New York, NY, USA},
volume = {37},
number = {7},
issn = {0001-0782},
url = {https://doi.org/10.1145/176789.176792},
doi = {10.1145/176789.176792},
journal = {Commun. ACM},
month = jul,
pages = {30–40},
numpages = {11}
}

@article{shneiderman1997direct,
author = {Shneiderman, Ben and Maes, Pattie},
title = {Direct manipulation vs. interface agents},
year = {1997},
issue_date = {Nov./Dec. 1997},
publisher = {Association for Computing Machinery},
address = {New York, NY, USA},
volume = {4},
number = {6},
issn = {1072-5520},
url = {https://doi.org/10.1145/267505.267514},
doi = {10.1145/267505.267514},
journal = {Interactions},
month = nov,
pages = {42–61},
numpages = {20}
}

@article{salaudeen2025measurement,
  title={Measurement to Meaning: A Validity-Centered Framework for AI Evaluation},
  author={Salaudeen, Olawale and Reuel, Anka and Ahmed, Ahmed and Bedi, Suhana and Robertson, Zachary and Sundar, Sudharsan and Domingue, Ben and Wang, Angelina and Koyejo, Sanmi},
  journal={arXiv preprint arXiv:2505.10573},
  year={2025}
}

@article{wallach2025position,
  title={Position: Evaluating generative ai systems is a social science measurement challenge},
  author={Wallach, Hanna and Desai, Meera and Cooper, A Feder and Wang, Angelina and Atalla, Chad and Barocas, Solon and Blodgett, Su Lin and Chouldechova, Alexandra and Corvi, Emily and Dow, P Alex and others},
  journal={arXiv preprint arXiv:2502.00561},
  year={2025}
}

@article{terry2023interactive,
  title={Interactive AI alignment: Specification, process, and evaluation alignment},
  author={Terry, Michael and Kulkarni, Chinmay and Wattenberg, Martin and Dixon, Lucas and Morris, Meredith Ringel},
  journal={arXiv preprint arXiv:2311.00710},
  year={2023}
}

@article{cemri2025multi,
  title={Why do multi-agent llm systems fail?},
  author={Cemri, Mert and Pan, Melissa Z and Yang, Shuyi and Agrawal, Lakshya A and Chopra, Bhavya and Tiwari, Rishabh and Keutzer, Kurt and Parameswaran, Aditya and Klein, Dan and Ramchandran, Kannan and others},
  journal={arXiv preprint arXiv:2503.13657},
  year={2025}
}

@inproceedings{parameswaran2024revisiting,
  author       = {Aditya G. Parameswaran and
                  Shreya Shankar and
                  Parth Asawa and
                  Naman Jain and
                  Yujie Wang},
  title        = {Revisiting Prompt Engineering via Declarative Crowdsourcing},
  booktitle    = {14th Conference on Innovative Data Systems Research, {CIDR} 2024,
                  Chaminade, HI, USA, January 14-17, 2024},
  publisher    = {www.cidrdb.org},
  year         = {2024},
  url          = {https://www.cidrdb.org/cidr2024/papers/p67-parameswaran.pdf},
  bibsource    = {dblp computer science bibliography, https://dblp.org}
}

@inproceedings{shankar2024validates,
  title={Who validates the validators? aligning llm-assisted evaluation of llm outputs with human preferences},
  author={Shankar, Shreya and Zamfirescu-Pereira, JD and Hartmann, Bj{\"o}rn and Parameswaran, Aditya and Arawjo, Ian},
  booktitle={Proceedings of the 37th Annual ACM Symposium on User Interface Software and Technology},
  pages={1--14},
  year={2024}
}

@article{pang2025interactive,
  title={Interactive Reasoning: Visualizing and Controlling Chain-of-Thought Reasoning in Large Language Models},
  author={Pang, Rock Yuren and Feng, KJ and Feng, Shangbin and Li, Chu and Shi, Weijia and Tsvetkov, Yulia and Heer, Jeffrey and Reinecke, Katharina},
  journal={arXiv preprint arXiv:2506.23678},
  year={2025}
}

@inproceedings{kazemitabaar2024improving,
  title={Improving steering and verification in AI-assisted data analysis with interactive task decomposition},
  author={Kazemitabaar, Majeed and Williams, Jack and Drosos, Ian and Grossman, Tovi and Henley, Austin Zachary and Negreanu, Carina and Sarkar, Advait},
  booktitle={Proceedings of the 37th Annual ACM Symposium on User Interface Software and Technology},
  pages={1--19},
  year={2024}
}

@misc{manus2025agent,
  author       = {Manus AI},
  title        = {Manus: General AI Agent that Bridges Mind and Action},
  year         = {2025},
  howpublished = {\url{https://manus.im/?index=1}}
}

@misc{anthropic2024introducing,
  author       = {Anthropic},
  title        = {Introducing computer use, a new Claude 3.5 Sonnet, and Claude 3.5 Haiku},
  year         = {2024},
  howpublished = {\url{https://www.anthropic.com/news/3-5-models-and-computer-use}}
}

@misc{anthropic2024claudecode,
  author       = {Anthropic},
  title        = {Introduction to agentic coding},
  year         = {2025},
  howpublished = {\url{https://claude.com/blog/introduction-to-agentic-coding}}
}

@misc{githubcopilot,
      author = {{GitHub}},
      title = {GitHub Copilot},
      year = {2021},
      url = {https://github.com/features/copilot},
}

@misc{openai2025deepresearch,
      author = {{OpenAI}},
      title = {Introducing deep research },
      year = {2025},
      url = {https://openai.com/index/introducing-deep-research/},
}

@misc{openai2025gpt51thinking,
      author = {{OpenAI}},
      title = {GPT-5.1: A smarter, more conversational ChatGPT},
      year = {2025},
      url = {https://openai.com/index/gpt-5-1/},
}

@misc{google2025deepresearch,
      author = {{Google}},
      title = {Gemini Deep Research},
      year = {2025},
      url = {https://gemini.google/overview/deep-research/},
}

@misc{aiagentsdirectory2025ai,
  author       = {{AI Agents Directory}},
  title        = {AI Agent Marketplace \& Directory | Find Top AI Agents \& AI Agent Solutions},
  year         = {2025},
  howpublished = {\url{https://aiagentsdirectory.com/}}
}

@misc{yao2024tau,
      title={$\tau$-bench: A Benchmark for Tool-Agent-User Interaction in Real-World Domains}, 
      author={Shunyu Yao and Noah Shinn and Pedram Razavi and Karthik Narasimhan},
      year={2024},
      eprint={2406.12045},
      archivePrefix={arXiv},
      primaryClass={cs.AI},
      url={https://arxiv.org/abs/2406.12045}, 
}

@misc{barres2025tau2,
      title={$\tau^2$-Bench: Evaluating Conversational Agents in a Dual-Control Environment}, 
      author={Victor Barres and Honghua Dong and Soham Ray and Xujie Si and Karthik Narasimhan},
      year={2025},
      eprint={2506.07982},
      archivePrefix={arXiv},
      primaryClass={cs.AI},
      url={https://arxiv.org/abs/2506.07982}, 
}

@article{drouin2024workarena,
  title={Workarena: How capable are web agents at solving common knowledge work tasks?},
  author={Drouin, Alexandre and Gasse, Maxime and Caccia, Massimo and Laradji, Issam H and Del Verme, Manuel and Marty, Tom and Boisvert, L{\'e}o and Thakkar, Megh and Cappart, Quentin and Vazquez, David and others},
  journal={arXiv preprint arXiv:2403.07718},
  year={2024}
}

@article{zhou2023webarena,
  title={Webarena: A realistic web environment for building autonomous agents},
  author={Zhou, Shuyan and Xu, Frank F and Zhu, Hao and Zhou, Xuhui and Lo, Robert and Sridhar, Abishek and Cheng, Xianyi and Ou, Tianyue and Bisk, Yonatan and Fried, Daniel and others},
  journal={arXiv preprint arXiv:2307.13854},
  year={2023}
}

@article{yoran2024assistantbench,
  title={Assistantbench: Can web agents solve realistic and time-consuming tasks?},
  author={Yoran, Ori and Amouyal, Samuel Joseph and Malaviya, Chaitanya and Bogin, Ben and Press, Ofir and Berant, Jonathan},
  journal={arXiv preprint arXiv:2407.15711},
  year={2024}
}

@article{xie2024osworld,
  title={Osworld: Benchmarking multimodal agents for open-ended tasks in real computer environments},
  author={Xie, Tianbao and Zhang, Danyang and Chen, Jixuan and Li, Xiaochuan and Zhao, Siheng and Cao, Ruisheng and Hua, Toh J and Cheng, Zhoujun and Shin, Dongchan and Lei, Fangyu and others},
  journal={Advances in Neural Information Processing Systems},
  volume={37},
  pages={52040--52094},
  year={2024}
}

@article{jimenez2023swe,
  title={Swe-bench: Can language models resolve real-world github issues?},
  author={Jimenez, Carlos E and Yang, John and Wettig, Alexander and Yao, Shunyu and Pei, Kexin and Press, Ofir and Narasimhan, Karthik},
  journal={arXiv preprint arXiv:2310.06770},
  year={2023}
}

@inproceedings{mialon2023gaia,
  title={Gaia: a benchmark for general ai assistants},
  author={Mialon, Gr{\'e}goire and Fourrier, Cl{\'e}mentine and Wolf, Thomas and LeCun, Yann and Scialom, Thomas},
  booktitle={The Twelfth International Conference on Learning Representations},
  year={2023}
}

@article{liu2024visualwebbench,
  title={Visualwebbench: How far have multimodal llms evolved in web page understanding and grounding?},
  author={Liu, Junpeng and Song, Yifan and Lin, Bill Yuchen and Lam, Wai and Neubig, Graham and Li, Yuanzhi and Yue, Xiang},
  journal={arXiv preprint arXiv:2404.05955},
  year={2024}
}

@article{yao2022webshop,
  title={Webshop: Towards scalable real-world web interaction with grounded language agents},
  author={Yao, Shunyu and Chen, Howard and Yang, John and Narasimhan, Karthik},
  journal={Advances in Neural Information Processing Systems},
  volume={35},
  pages={20744--20757},
  year={2022}
}

@inproceedings{shi2017world,
  title={World of bits: An open-domain platform for web-based agents},
  author={Shi, Tianlin and Karpathy, Andrej and Fan, Linxi and Hernandez, Jonathan and Liang, Percy},
  booktitle={International Conference on Machine Learning},
  pages={3135--3144},
  year={2017},
  organization={PMLR}
}

@article{deng2023mind2web,
  title={Mind2web: Towards a generalist agent for the web},
  author={Deng, Xiang and Gu, Yu and Zheng, Boyuan and Chen, Shijie and Stevens, Sam and Wang, Boshi and Sun, Huan and Su, Yu},
  journal={Advances in Neural Information Processing Systems},
  volume={36},
  pages={28091--28114},
  year={2023}
}

@article{pan2024webcanvas,
  title={Webcanvas: Benchmarking web agents in online environments},
  author={Pan, Yichen and Kong, Dehan and Zhou, Sida and Cui, Cheng and Leng, Yifei and Jiang, Bing and Liu, Hangyu and Shang, Yanyi and Zhou, Shuyan and Wu, Tongshuang and others},
  journal={arXiv preprint arXiv:2406.12373},
  year={2024}
}

@article{buccinca2021trust,
  title={To trust or to think: cognitive forcing functions can reduce overreliance on AI in AI-assisted decision-making},
  author={Bu{\c{c}}inca, Zana and Malaya, Maja Barbara and Gajos, Krzysztof Z},
  journal={Proceedings of the ACM on Human-computer Interaction},
  volume={5},
  number={CSCW1},
  pages={1--21},
  year={2021},
  publisher={ACM New York, NY, USA}
}

@article{miller2019explanation,
  title={Explanation in artificial intelligence: Insights from the social sciences},
  author={Miller, Tim},
  journal={Artificial intelligence},
  volume={267},
  pages={1--38},
  year={2019},
  publisher={Elsevier}
}

@inproceedings{amershi2019guidelines,
author = {Amershi, Saleema and Weld, Dan and Vorvoreanu, Mihaela and Fourney, Adam and Nushi, Besmira and Collisson, Penny and Suh, Jina and Iqbal, Shamsi and Bennett, Paul N. and Inkpen, Kori and Teevan, Jaime and Kikin-Gil, Ruth and Horvitz, Eric},
title = {Guidelines for Human-AI Interaction},
year = {2019},
isbn = {9781450359702},
publisher = {Association for Computing Machinery},
address = {New York, NY, USA},
url = {https://doi.org/10.1145/3290605.3300233},
doi = {10.1145/3290605.3300233},
booktitle = {Proceedings of the 2019 CHI Conference on Human Factors in Computing Systems},
pages = {1–13},
numpages = {13},
location = {Glasgow, Scotland Uk},
series = {CHI '19}
}

@article{bansal2024challenges,
  title={Challenges in human-agent communication},
  author={Bansal, Gagan and Vaughan, Jennifer Wortman and Amershi, Saleema and Horvitz, Eric and Fourney, Adam and Mozannar, Hussein and Dibia, Victor and Weld, Daniel S},
  journal={arXiv preprint arXiv:2412.10380},
  year={2024}
}

@inproceedings{kim2024im,
  title={" I'm Not Sure, But...": Examining the Impact of Large Language Models' Uncertainty Expression on User Reliance and Trust},
  author={Kim, Sunnie SY and Liao, Q Vera and Vorvoreanu, Mihaela and Ballard, Stephanie and Vaughan, Jennifer Wortman},
  booktitle={Proceedings of the 2024 ACM conference on fairness, accountability, and transparency},
  pages={822--835},
  year={2024}
}

@inproceedings{bo2025rely,
author = {Bo, Jessica Y and Wan, Sophia and Anderson, Ashton},
title = {To Rely or Not to Rely? Evaluating Interventions for Appropriate Reliance on Large Language Models},
year = {2025},
isbn = {9798400713941},
publisher = {Association for Computing Machinery},
address = {New York, NY, USA},
url = {https://doi.org/10.1145/3706598.3714097},
doi = {10.1145/3706598.3714097},
booktitle = {Proceedings of the 2025 CHI Conference on Human Factors in Computing Systems},
articleno = {905},
numpages = {23},
location = {
},
series = {CHI '25}
}

@article{vaccaro2024combinations,
  title={When combinations of humans and AI are useful: A systematic review and meta-analysis},
  author={Vaccaro, Michelle and Almaatouq, Abdullah and Malone, Thomas},
  journal={Nature Human Behaviour},
  volume={8},
  number={12},
  pages={2293--2303},
  year={2024},
  publisher={Nature Publishing Group UK London}
}

@article{vasconcelos2023explanations,
  title={Explanations can reduce overreliance on ai systems during decision-making},
  author={Vasconcelos, Helena and J{\"o}rke, Matthew and Grunde-McLaughlin, Madeleine and Gerstenberg, Tobias and Bernstein, Michael S and Krishna, Ranjay},
  journal={Proceedings of the ACM on Human-Computer Interaction},
  volume={7},
  number={CSCW1},
  pages={1--38},
  year={2023},
  publisher={ACM New York, NY, USA}
}

@inbook{he2025plan,
author = {He, Gaole and Demartini, Gianluca and Gadiraju, Ujwal},
title = {Plan-Then-Execute: An Empirical Study of User Trust and Team Performance When Using LLM Agents As A Daily Assistant},
year = {2025},
isbn = {9798400713941},
publisher = {Association for Computing Machinery},
address = {New York, NY, USA},
url = {https://doi.org/10.1145/3706598.3713218},
booktitle = {Proceedings of the 2025 CHI Conference on Human Factors in Computing Systems},
articleno = {414},
numpages = {22}
}

@article{chen2025toward,
  title={Toward a human-centered evaluation framework for trustworthy llm-powered gui agents},
  author={Chen, Chaoran and Zhang, Zhiping and Khalilov, Ibrahim and Guo, Bingcan and Gebreegziabher, Simret A and Ye, Yanfang and Xiao, Ziang and Yao, Yaxing and Li, Tianshi and Li, Toby Jia-Jun},
  journal={arXiv preprint arXiv:2504.17934},
  year={2025}
}

@article{grunde2025designing,
  title={Designing LLM chains by adapting techniques from crowdsourcing workflows},
  author={Grunde-McLaughlin, Madeleine and Lam, Michelle S and Krishna, Ranjay and Weld, Daniel S and Heer, Jeffrey},
  journal={ACM Transactions on Computer-Human Interaction},
  volume={32},
  number={3},
  pages={1--57},
  year={2025},
  publisher={ACM New York, NY}
}

@inproceedings{wu2022ai,
author = {Wu, Tongshuang and Terry, Michael and Cai, Carrie Jun},
title = {AI Chains: Transparent and Controllable Human-AI Interaction by Chaining Large Language Model Prompts},
year = {2022},
isbn = {9781450391573},
publisher = {Association for Computing Machinery},
address = {New York, NY, USA},
url = {https://doi.org/10.1145/3491102.3517582},
doi = {10.1145/3491102.3517582},
booktitle = {Proceedings of the 2022 CHI Conference on Human Factors in Computing Systems},
articleno = {385},
numpages = {22},
location = {New Orleans, LA, USA},
series = {CHI '22}
}

@inproceedings{wu2022promptchainer,
author = {Wu, Tongshuang and Jiang, Ellen and Donsbach, Aaron and Gray, Jeff and Molina, Alejandra and Terry, Michael and Cai, Carrie J},
title = {PromptChainer: Chaining Large Language Model Prompts through Visual Programming},
year = {2022},
isbn = {9781450391566},
publisher = {Association for Computing Machinery},
address = {New York, NY, USA},
url = {https://doi.org/10.1145/3491101.3519729},
doi = {10.1145/3491101.3519729},
booktitle = {Extended Abstracts of the 2022 CHI Conference on Human Factors in Computing Systems},
articleno = {359},
numpages = {10},
location = {New Orleans, LA, USA},
series = {CHI EA '22}
}

@inproceedings{epperson2025interactive,
author = {Epperson, Will and Bansal, Gagan and Dibia, Victor C and Fourney, Adam and Gerrits, Jack and Zhu, Erkang (Eric) and Amershi, Saleema},
title = {Interactive Debugging and Steering of Multi-Agent AI Systems},
year = {2025},
isbn = {9798400713941},
publisher = {Association for Computing Machinery},
address = {New York, NY, USA},
url = {https://doi.org/10.1145/3706598.3713581},
doi = {10.1145/3706598.3713581},
booktitle = {Proceedings of the 2025 CHI Conference on Human Factors in Computing Systems},
articleno = {156},
numpages = {15},
location = {
},
series = {CHI '25}
}

@article{pan2025agentcoord,
  title={Agentcoord: Visually exploring coordination strategy for llm-based multi-agent collaboration},
  author={Pan, Bo and Lu, Jiaying and Wang, Ke and Zheng, Li and Wen, Zhen and Feng, Yingchaojie and Zhu, Minfeng and Chen, Wei},
  journal={Computers \& graphics},
  pages={104338},
  year={2025},
  publisher={Elsevier}
}

@article{liao2021human,
  title={Human-centered explainable ai (xai): From algorithms to user experiences},
  author={Liao, Q Vera and Varshney, Kush R},
  journal={arXiv preprint arXiv:2110.10790},
  year={2021}
}

@article{goulet2018review,
author = {Goulet-Pelletier, Jean-Christophe and Cousineau, Denis},
year = {2018},
month = {12},
pages = {242-265},
title = {A review of effect sizes and their confidence intervals, Part {I}: The Cohen's d family},
volume = {14},
journal = {The Quantitative Methods for Psychology},
doi = {10.20982/tqmp.14.4.p242}
}

@incollection{dragicevic2016fair,
  title={Fair statistical communication in HCI},
  author={Dragicevic, Pierre},
  booktitle={Modern statistical methods for HCI},
  pages={291--330},
  year={2016},
  publisher={Springer}
}

@article{braun2006using,
  title={Using thematic analysis in psychology},
  author={Braun, Virginia and Clarke, Victoria},
  journal={Qualitative research in psychology},
  volume={3},
  number={2},
  pages={77--101},
  year={2006},
  publisher={Taylor \& Francis}
}

@article{agarwal2024faithfulness,
  title={Faithfulness vs. plausibility: On the (un) reliability of explanations from large language models},
  author={Agarwal, Chirag and Tanneru, Sree Harsha and Lakkaraju, Himabindu},
  journal={arXiv preprint arXiv:2402.04614},
  year={2024}
}

@inproceedings{turpin2023language,
author = {Turpin, Miles and Michael, Julian and Perez, Ethan and Bowman, Samuel R.},
title = {Language models don't always say what they think: unfaithful explanations in chain-of-thought prompting},
year = {2023},
publisher = {Curran Associates Inc.},
address = {Red Hook, NY, USA},
booktitle = {Proceedings of the 37th International Conference on Neural Information Processing Systems},
articleno = {3275},
numpages = {14},
location = {New Orleans, LA, USA},
series = {NeuIPS '23}
}

@inproceedings{madsen2024are,
    title = "Are self-explanations from Large Language Models faithful?",
    author = "Madsen, Andreas  and
      Chandar, Sarath  and
      Reddy, Siva",
    editor = "Ku, Lun-Wei  and
      Martins, Andre  and
      Srikumar, Vivek",
    booktitle = "Findings of the Association for Computational Linguistics: ACL 2024",
    month = aug,
    year = "2024",
    address = "Bangkok, Thailand",
    publisher = "Association for Computational Linguistics",
    url = "https://aclanthology.org/2024.findings-acl.19/",
    doi = "10.18653/v1/2024.findings-acl.19",
    pages = "295--337"
}

@inproceedings{wu2024autogen,
  title={Autogen: Enabling next-gen LLM applications via multi-agent conversations},
  author={Wu, Qingyun and Bansal, Gagan and Zhang, Jieyu and Wu, Yiran and Li, Beibin and Zhu, Erkang and Jiang, Li and Zhang, Xiaoyun and Zhang, Shaokun and Liu, Jiale and others},
  booktitle={First Conference on Language Modeling},
  year={2024}
}

@inproceedings{du2024improving,
author = {Du, Yilun and Li, Shuang and Torralba, Antonio and Tenenbaum, Joshua B. and Mordatch, Igor},
title = {Improving factuality and reasoning in language models through multiagent debate},
year = {2024},
publisher = {JMLR.org},
booktitle = {Proceedings of the 41st International Conference on Machine Learning},
articleno = {467},
numpages = {31},
location = {Vienna, Austria},
series = {ICML'24}
}

@inproceedings{huffaker2020crowdsourced,
author = {Huffaker, Jordan S. and Kummerfeld, Jonathan K. and Lasecki, Walter S. and Ackerman, Mark S.},
title = {Crowdsourced Detection of Emotionally Manipulative Language},
year = {2020},
isbn = {9781450367080},
publisher = {Association for Computing Machinery},
address = {New York, NY, USA},
url = {https://doi.org/10.1145/3313831.3376375},
doi = {10.1145/3313831.3376375},
booktitle = {Proceedings of the 2020 CHI Conference on Human Factors in Computing Systems},
pages = {1–14},
numpages = {14},
location = {Honolulu, HI, USA},
series = {CHI '20}
}

@article{liem2011iterative,
  title={An iterative dual pathway structure for speech-to-text transcription},
  author={Liem, Beatrice and Zhang, Haoqi and Chen, Yiling},
  year={2011},
  publisher={Association for the Advancement of Artificial Intelligence}
}

@inproceedings{chen2019cicero,
author = {Chen, Quanze and Bragg, Jonathan and Chilton, Lydia B. and Weld, Dan S.},
title = {Cicero: Multi-Turn, Contextual Argumentation for Accurate Crowdsourcing},
year = {2019},
isbn = {9781450359702},
publisher = {Association for Computing Machinery},
address = {New York, NY, USA},
url = {https://doi.org/10.1145/3290605.3300761},
doi = {10.1145/3290605.3300761},
booktitle = {Proceedings of the 2019 CHI Conference on Human Factors in Computing Systems},
pages = {1–14},
numpages = {14},
location = {Glasgow, Scotland Uk},
series = {CHI '19}
}

@inproceedings{kobayashi2018empirical,
  title={An empirical study on short-and long-term effects of self-correction in crowdsourced microtasks},
  author={Kobayashi, Masaki and Morita, Hiromi and Matsubara, Masaki and Shimizu, Nobuyuki and Morishima, Atsuyuki},
  booktitle={Proceedings of the AAAI Conference on Human Computation and Crowdsourcing},
  volume={6},
  pages={79--87},
  year={2018}
}

@inproceedings{kittur2011crowdforge,
author = {Kittur, Aniket and Smus, Boris and Khamkar, Susheel and Kraut, Robert E.},
title = {CrowdForge: crowdsourcing complex work},
year = {2011},
isbn = {9781450307161},
publisher = {Association for Computing Machinery},
address = {New York, NY, USA},
url = {https://doi.org/10.1145/2047196.2047202},
doi = {10.1145/2047196.2047202},
booktitle = {Proceedings of the 24th Annual ACM Symposium on User Interface Software and Technology},
pages = {43–52},
numpages = {10},
location = {Santa Barbara, California, USA},
series = {UIST '11}
}

@inproceedings{bansal2021does,
  title={Does the whole exceed its parts? the effect of ai explanations on complementary team performance},
  author={Bansal, Gagan and Wu, Tongshuang and Zhou, Joyce and Fok, Raymond and Nushi, Besmira and Kamar, Ece and Ribeiro, Marco Tulio and Weld, Daniel},
  booktitle={Proceedings of the 2021 CHI conference on human factors in computing systems},
  pages={1--16},
  year={2021}
}

@article{vasconcelos2025generation,
author = {Vasconcelos, Helena and Bansal, Gagan and Fourney, Adam and Liao, Q. Vera and Wortman Vaughan, Jennifer},
title = {Generation Probabilities Are Not Enough: Uncertainty Highlighting in AI Code Completions},
year = {2025},
issue_date = {February 2025},
publisher = {Association for Computing Machinery},
address = {New York, NY, USA},
volume = {32},
number = {1},
issn = {1073-0516},
url = {https://doi.org/10.1145/3702320},
doi = {10.1145/3702320},
journal = {ACM Trans. Comput.-Hum. Interact.},
month = apr,
articleno = {4},
numpages = {30}
}

@article{wu2020managing,
  title={Managing uncertainty in AI-enabled decision making and achieving sustainability},
  author={Wu, Junyi and Shang, Shari},
  journal={Sustainability},
  volume={12},
  number={21},
  pages={8758},
  year={2020},
  publisher={MDPI}
}

@inproceedings{dibia2024autogenstudio,
    title = "{AUTOGEN} {STUDIO}: A No-Code Developer Tool for Building and Debugging Multi-Agent Systems",
    author = "Dibia, Victor  and
      Chen, Jingya  and
      Bansal, Gagan  and
      Syed, Suff  and
      Fourney, Adam  and
      Zhu, Erkang  and
      Wang, Chi  and
      Amershi, Saleema",
    editor = "Hernandez Farias, Delia Irazu  and
      Hope, Tom  and
      Li, Manling",
    booktitle = "Proceedings of the 2024 Conference on Empirical Methods in Natural Language Processing: System Demonstrations",
    month = nov,
    year = "2024",
    address = "Miami, Florida, USA",
    publisher = "Association for Computational Linguistics",
    url = "https://aclanthology.org/2024.emnlp-demo.8/",
    doi = "10.18653/v1/2024.emnlp-demo.8",
    pages = "72--79"
}

\newpage
\section{Appendix}
\begin{figure*}[t]
 \centering 
 \includegraphics[width=\linewidth]{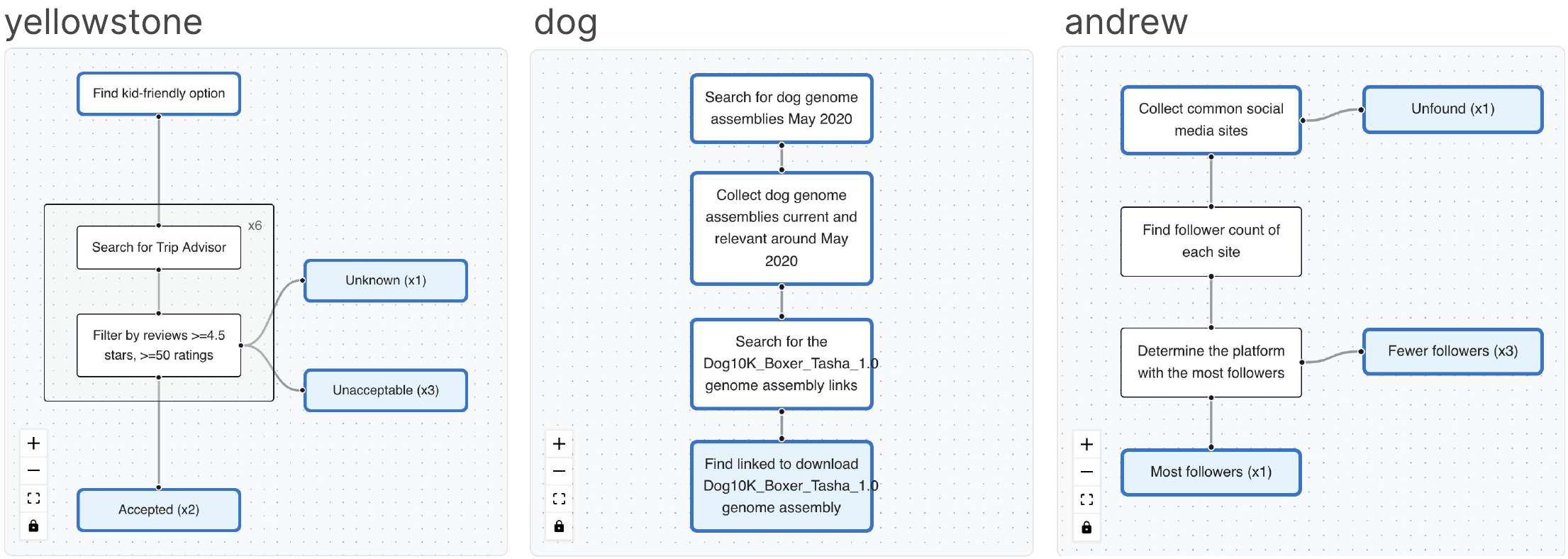}
 \caption{Flowchart tasks. Keywords relate to Table~\ref{tab:app-s2-tasks}.} 
 \Description{Three columns for three flowchart tasks. Each has nodes with action items, some of which are highlighted in blue. Outcome nodes have a light blue background.}
\label{fig:probe-flowchart} 
\end{figure*}

\begin{figure*}[t]
 \centering 
 \includegraphics[width=\linewidth]{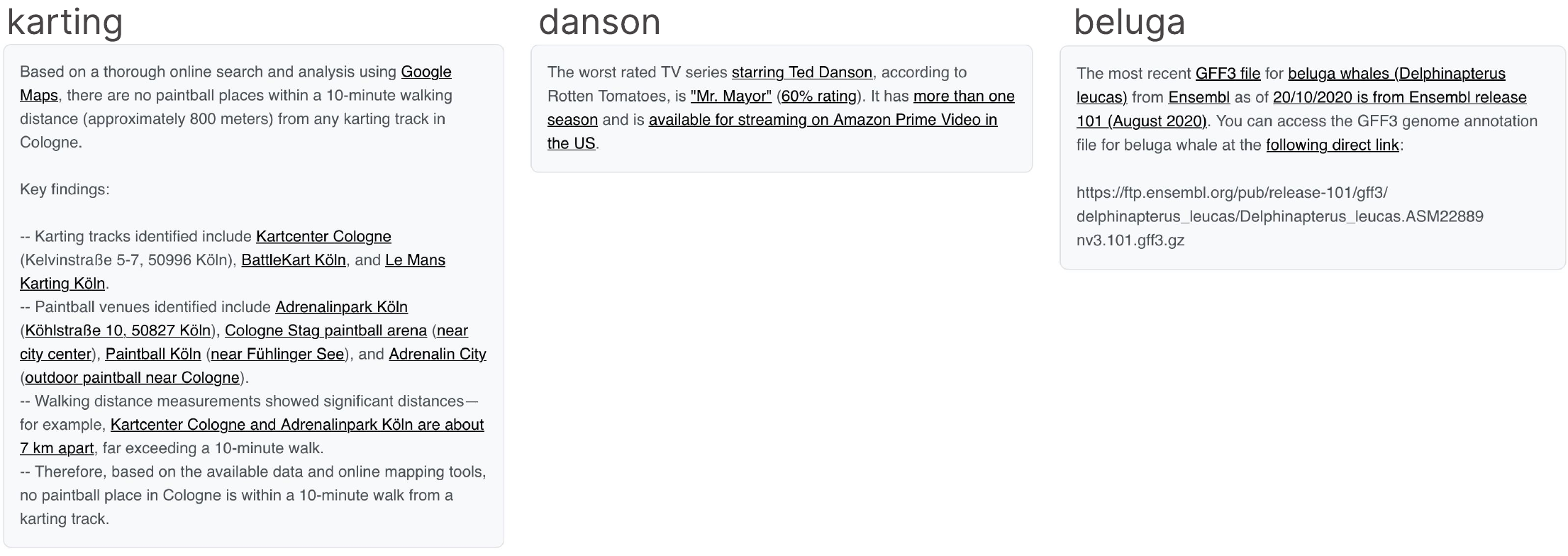}
 \caption{Citation tasks. Keywords relate to Table~\ref{tab:app-s2-tasks}.} 
 \Description{Three columns, each of which shows the citation description for the three citation tasks.}
\label{fig:probe-citation} 
\end{figure*}

\begin{figure*}[t]
 \centering 
 \includegraphics[width=\linewidth]{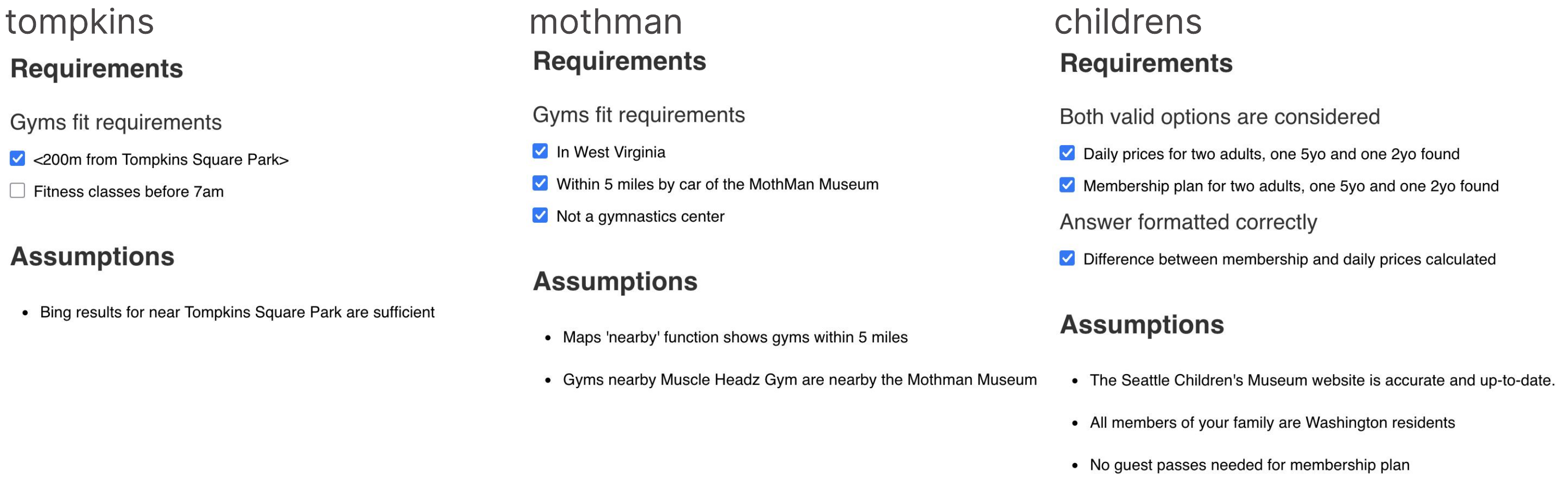}
 \caption{Specification tasks. Keywords relate to Table~\ref{tab:app-s2-tasks}.} 
 \Description{Three columns, each of which shows the specification for each of the specification tasks. }
\label{fig:probe-spec} 
\end{figure*}

\begin{figure*}[t]
 \centering 
 \includegraphics[width=\linewidth]{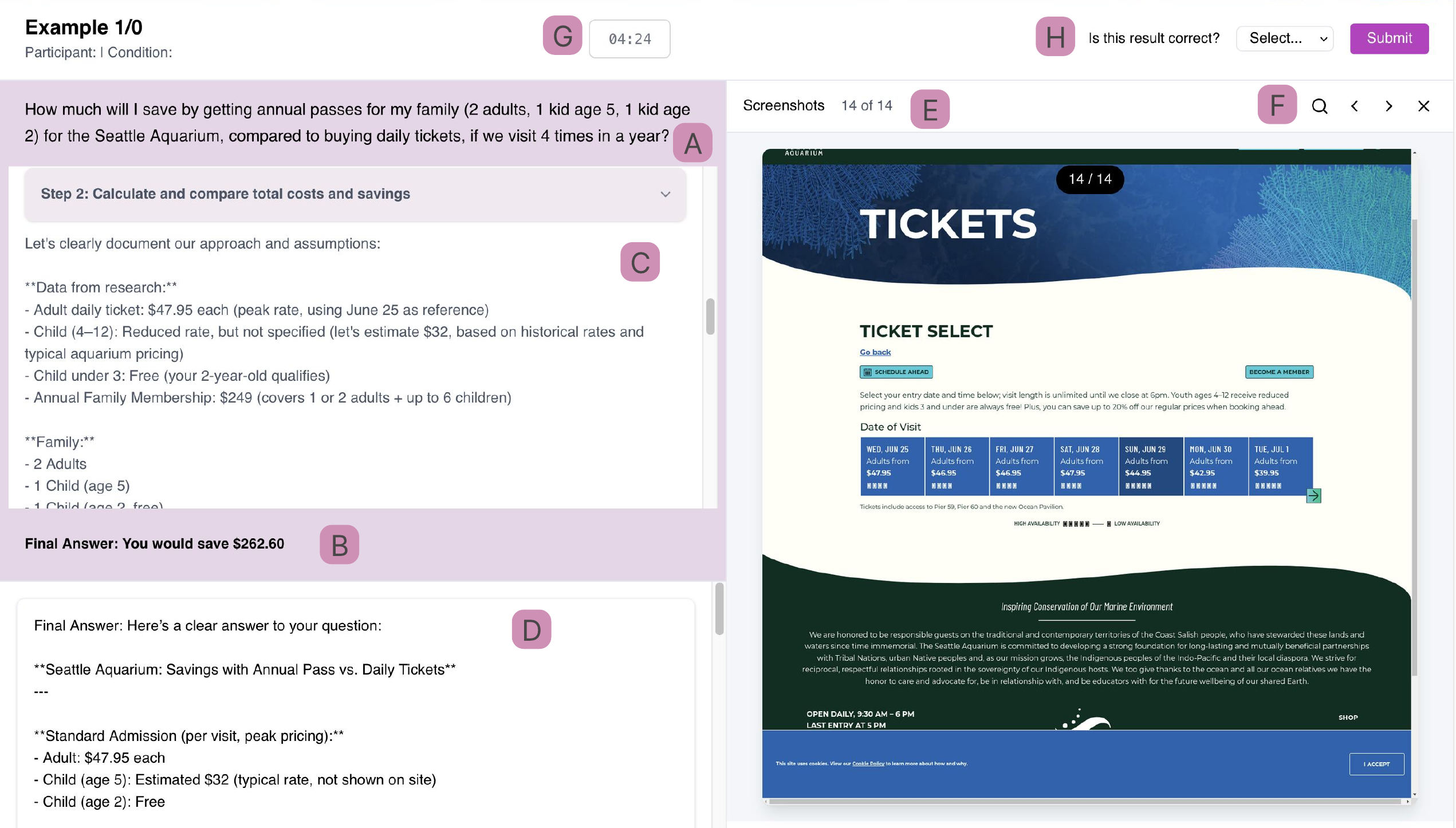}
 \caption{Control study baseline interface. The content is near-identical Magentic-UI's, but more visibly similar to the treatment condition. The display shows the task [A] and final answer [B], with a text description of actions [C] and a summarized final answer [D]. The screenshots are available [E], with one addition of a magnification feature [F]. Participants have five minutes [G] to determine if the output is correct [H].} 
 \Description{This figure shows an interface display. The top bar has a timer countdown and a question ``Is this result correct'' with a dropdown selection and a submit button. There is a pink box that contains the task text at the top left, then a white box inset that contains the trace of actions, and then a final answer is briefly stated. Below this pink box is a white box containing a ``Final Answer'' summary with more detail. The right side has a ``Screenshots'' header indicating we are on screenshot ``14 of 14''. There is an icon of a magnifying glass and arrow keys. Below this header is the screenshot itself.}
\label{fig:baseline-interface} 
\end{figure*}

From the formative study, we report the number of attempts with technical difficulties (Table~\ref{tab:app-form-errors}). From the design probe study, we report the tasks (Table~\ref{tab:app-s2-tasks}), their associated summarizations (Figures~\ref{fig:probe-flowchart}-\ref{fig:probe-spec}), participant preferences (Tables~\ref{tab:app-pref-overall} and~\ref{tab:app-pref-vali}), and results regarding estimated time to complete (Section~\ref{sec:app-estimation}). From the controlled study, we provide the tasks (Table~\ref{tab:app-s3-tasks}), confidence scale (Table~\ref{tab:app-s3conf}), and the baseline interface (Figure~\ref{fig:baseline-interface}).

\begin{table*}[h]
\caption{Outcome instances from the formative study.}
\label{tab:app-form-errors}
\centering
\resizebox{0.5\linewidth}{!}{

\begin{tabular}{lrrr}
                        & \multicolumn{1}{r}{Task 1} & \multicolumn{1}{r}{Task 2} & \multicolumn{1}{r}{Task 3} \\ \hline
Correct                 & 8                          & 1                          & 2                          \\
Acceptable              & \multicolumn{1}{r}{N/A}    & 4                          & \multicolumn{1}{r}{N/A}    \\
Incorrect               & 1                          & 5                          & 10                         \\
Incompete -- timing     & 2                          & 4                          & 0                          \\
Incomplete -- technical & 3                          & 1                          & 1                         
\end{tabular}
}
\end{table*}

\begin{table*}[h]
\caption{Tasks from the design probe study. The Insuff. column delineates if the answer ``Insufficient'' counts as correct.}
\label{tab:app-s2-tasks}
\centering
\resizebox{\linewidth}{!}{
\begin{tabular}{lllll}
Key         & Task                                                                                                                                                                                                                                             & Gound Truth Ans.                                                                                                                                 & Presented Ans.                                                                                                                                                          & Insuff. \\ \hline
\multicolumn{5}{l}{FLOWCHART}                                                                                                                                                                                                                                                                                                                                                                                                                                                                                                                                                 \\ \hline
yellowstone & \begin{tabular}[c]{@{}l@{}}What are hikes in Yellowstone that have been recommended \\ by at least three different people with kids and are highly rated \\ on TripAdvisor (an average from 4.5/5 from at least 50 reviews)?\end{tabular}        & \begin{tabular}[c]{@{}l@{}}Mystic Falls Trail\\ Fountain Paint Pot\\ Trout Lake Trail\\ Lone Star Geyser\end{tabular}                            & \begin{tabular}[c]{@{}l@{}}1. Fairy Falls Trail\\  2. Trout Lake Trail\end{tabular}                                                                                     & Yes     \\ \hline
dog         & \begin{tabular}[c]{@{}l@{}}The dog genome was first mapped in 2004 and has been updated \\ Several times since. What is the link to the files that were most \\ recent in May 2020?\end{tabular}                                                 & {\color[HTML]{467886} \begin{tabular}[c]{@{}l@{}}https://ftp.ensembl.org/pub/\\ release-100/fasta/\\ canis\_lupus\_familiaris/dna/\end{tabular}} & \begin{tabular}[c]{@{}l@{}}https://www.ncbi.nlm.nih.gov\\ /datasets/genome/\\ GCF\_000002285.5/\end{tabular}                                                            & No      \\ \hline
andew       & \begin{tabular}[c]{@{}l@{}}On which social media platform does Andrew Ng have the \\ most followers?\end{tabular}                                                                                                                                & Linked-In                                                                                                                                        & Linked-In                                                                                                                                                               & Yes     \\ \hline
\multicolumn{5}{l}{CITATION}                                                                                                                                                                                                                                                                                                                                                                                                                                                                                                                                                  \\ \hline
karting     & \begin{tabular}[c]{@{}l@{}}Which paintball places in Cologne, Germany are within a \\ 10 minute walk from a karting track?\end{tabular}                                                                                                          & Adrenalinpark Köln                                                                                                                               & None available                                                                                                                                                          & Yes     \\ \hline
danson      & \begin{tabular}[c]{@{}l@{}}What is the worst rated series (according to Rotten Tomatoes) \\ with more than 1 season that Ted Danson has starred in and is \\ available on Amazon Prime Video (US)?\end{tabular}                                  & CSI: Cyber                                                                                                                                       & Mr. Mayor                                                                                                                                                               & Yes     \\ \hline
beluga      & \begin{tabular}[c]{@{}l@{}}What is the link to the GFF3 file from Ensembl for beluga \\ whales that was the most recent one on 20/10/2020?\end{tabular}                                                                                          & {\color[HTML]{467886} \begin{tabular}[c]{@{}l@{}}https://ftp.ensembl.org/pub/\\ release-101/gff3/\\ delphinapterus\_leucas/\end{tabular}}        & \begin{tabular}[c]{@{}l@{}}https://ftp.ensembl.org/pub/release-101/\\ gff3/delphinapterus\_leucas/\\ Delphinapterus\_leucas.ASM22889 \\ nv3.101.gff3.gz\end{tabular}    & No      \\ \hline
\multicolumn{5}{l}{SPECIFICATION}                                                                                                                                                                                                                                                                                                                                                                                                                                                                                                                                             \\ \hline
tompkins    & \begin{tabular}[c]{@{}l@{}}What gyms are located within 200 meters of Tompkins\\ Square Park that offer fitness classes before 7am?\end{tabular}                                                                                                 & \begin{tabular}[c]{@{}l@{}}Avea Pilates\\ CrossFit East River\\ TPML Fitness\end{tabular}                                                        & No gyms available                                                                                                                                                       & Yes     \\ \hline
mothman     & \begin{tabular}[c]{@{}l@{}}Which gyms (not including gymnastics centers) in West \\ Virginia are within 5 miles (by car) of the Mothman Museum?\end{tabular}                                                                                     & \begin{tabular}[c]{@{}l@{}}The Root Sports \& Fitness Center\\  Muscle Headz Gym\end{tabular}                                                    & \begin{tabular}[c]{@{}l@{}}Muscle Headz Gym\\  The Root Sports \& Fitness Center\\  Fit Culture 24/7\\  The Warehouse Fitness Center LLC\\  Uplift Fitness\end{tabular} & No      \\ \hline
childrens   & \begin{tabular}[c]{@{}l@{}}How much will I save by getting annual passes for my family (2 \\ adults, 1 kid age 5, 1 kid age 2) for the Seattle Children Museum, \\ compared to buying daily tickets, if we visit 4 times in a year?\end{tabular} & \$61                                                                                                                                             & \$61                                                                                                                                                                    & No     
\end{tabular}
}
\end{table*}

\begin{table*}[h]
\caption{Overall preference rankings from the design probes study. Manually ordered by similarity.}
\label{tab:app-pref-overall}
\centering
\resizebox{0.7\linewidth}{!}{
\begin{tabular}{lllll}
Participant & Rank 1 (high) & Rank 2      & Rank 3      & Rank 4      \\ \hline
P6          & Flowchart     & Spec        & Citation    & Magentic-UI \\
P3          & Flowchart     & Spec        & Magentic-UI & Citation    \\
P1          & Spec          & Flowchart   & Citation    & Magentic-UI \\
P5          & Spec          & Flowchart   & Citation    & Magentic-UI \\
P4          & Spec          & Citation    & Flowchart   & Magentic-UI \\
P8          & Spec          & Citation    & Flowchart   & Magentic-UI \\
P9          & Spec          & Magentic-UI & Citation    & Flowchart   \\
P7          & Citation      & Spec        & Magentic-UI & Flowchart   \\
P12         & Citation      & Magentic-UI & Spec        & Flowchart   \\
P11         & Magentic-UI   & Citation    & Spec        & Flowchart   \\
P10         & Citation      & Magentic-UI & Flowchart   & Spec        \\
P2          & Citation      & Flowchart   & Magentic-UI & Spec       
\end{tabular}
}
\end{table*}

\begin{table*}[h]
\caption{Expected validation support rankings from the design probes study. Manually ordered by similarity. Only asked to P5-P12}
\label{tab:app-pref-vali}
\centering
\resizebox{0.7\linewidth}{!}{

\begin{tabular}{lllll}
Participant & Rank 1 (high) & Rank 2      & Rank 3      & Rank 4      \\ \hline
P7          & Citation      & Magentic-UI & Spec        & Flowchart   \\
P10         & Citation      & Magentic-UI & Flowchart   & Spec        \\
P12         & Magentic-UI   & Citation    & Flowchart   & Spec        \\
P5          & Magentic-UI   & Spec        & Flowchart   & Citation    \\
P6          & Magentic-UI   & Spec        & Flowchart   & Citation    \\
P8          & Spec          & Flowchart   & Citation    & Magentic-UI \\
P9          & Spec          & Citation    & Magentic-UI & Flowchart   \\
P11         & Spec          & Magentic-UI & Citation    & Flowchart  
\end{tabular}
}
\end{table*}

\subsection{Estimated time in design probes study}
\label{sec:app-estimation}
At the beginning of each task, the interface asked participants to estimate the amount of time they expected the task to take them to complete on their own. 
We found a negligible correlation between the expected time and the time taken to review (Spearman's rho of -0.17, p=0.0828). 
Additionally, the mean estimated time for correctly answered (mean=7.83, std=2.38) and incorrectly answered (mean=7.20, std=2.71) questions were well within a standard deviation. 
Finally, participants did not show a substantial difference in estimated time taken for tasks across the Spec (mean=7.09, std=2.88), Citation (mean=7.66, std=2.48), and Flowchart (mean=7.51, std=2.49) methods.

\begin{table*}[h]
\caption{Tasks from the control study.}
\label{tab:app-s3-tasks}
\centering
\resizebox{\linewidth}{!}{
\begin{tabular}{lll}
Task                                                                                                                                                                                                                                                              & Ground Truth Ans.                                                                                                         & Presented Ans.                                                                                                            \\ \hline
 
SET 1                                                                                                                                                                                                                                                             &                                                                                                                           &                                                                                                                           \\ \hline
\begin{tabular}[c]{@{}l@{}}Which Fidelity international emerging markets equity mutual fund \\ with \$0 transaction fees had the lowest percentage increase \\ between May 2019 to May 2024?\end{tabular}                                                         & FPADX                                                                                                                     & FEDDX                                                                                                                     \\ \hline
\begin{tabular}[c]{@{}l@{}}What is the highest rated (according to IMDB) Isabelle Adjani feature \\ film that is less than 2 hours and is available on Vudu (now called \\ Fandango at Home) to buy or rent?\end{tabular}                                         & Nosferatu the Vampyre                                                                                                     & No film fit the criteria.                                                                                                 \\ \hline
\begin{tabular}[c]{@{}l@{}}How much will I save by getting annual passes for my family (2 adults, \\ 1 kid age 5, 1 kid age 2) for the Seattle Children Museum, compared to \\ buying daily tickets, if we visit 4 times in a year?\end{tabular}                  & \$61                                                                                                                      & \$61                                                                                                                      \\ \hline
\begin{tabular}[c]{@{}l@{}}Which members of Fubo's Management Team joined the company \\ during the same year Fubo's IPO happened?\end{tabular}                                                                                                                   & Gina DiGioa                                                                                                               & Gina DiGoia                                                                                                               \\ \hline

SET 2                                                                                                                                                                                                                                                             &                                                                                                                           &                                                                                                                           \\ \hline
\begin{tabular}[c]{@{}l@{}}Which members of Apple’s Board of Directors did not hold C-suite \\ positions at their companies when they joined the board?\end{tabular}                                                                                              & \begin{tabular}[c]{@{}l@{}}Wanda Austin\\ Ronald D. Sugar\\ Sue Wagner\end{tabular}                                       & Susan L. Wagner                                                                                                           \\ \hline
\begin{tabular}[c]{@{}l@{}}How much will I save by getting annual passes for a group of 4 adults \\ and 1 student for the Philadelphia Museum of Art, compared to buying \\ daily tickets, if we visit 5 times in a year (all non-consecutive days)?\end{tabular} & \$265                                                                                                                     & \$195                                                                                                                     \\ \hline
\begin{tabular}[c]{@{}l@{}}The dog genome has been updated several times. What is the link to \\ the files that were most recent in May 2020?\end{tabular}                                                                                                        & \begin{tabular}[c]{@{}l@{}}https://ftp.ensembl.org/pub/\\ release-100/fasta/\\ canis\_lupus\_familiaris/dna/\end{tabular} & \begin{tabular}[c]{@{}l@{}}https://ftp.ensembl.org/pub/\\ release-100/fasta/\\ canis\_lupus\_familiaris/dna/\end{tabular} \\ \hline
\begin{tabular}[c]{@{}l@{}}What is the worst rated series (according to Rotten Tomatoes) with \\ more than 1 season that Ted Danson has starred in and is available on \\ Amazon Prime Video (US)?\end{tabular}                                                   & CSI: Cyber                                                                                                                & CSI: Cyber                                                                                                               
\end{tabular}
}
\end{table*}

\begin{table*}[t]
\caption{Confidence scale. The scale values participants selected from for the question ``How confident are you in your previous answer?''}
\label{tab:app-s3conf}
\centering
\resizebox{0.3\linewidth}{!}{
\begin{tabular}{ll}
Value & Descriptor           \\ \hline
1           & Not at all confident \\
2           & Slightly confident   \\
3           & Somewhat confident   \\
4           & Moderately confident \\
5           & Quite confident      \\
6           & Very confident       \\
7           & Extremely confident 
\end{tabular}
}
\end{table*}

\end{document}